\documentclass[aps,prd,preprint]{revtex4-1}
\usepackage{amsmath, bm, natbib, latexsym}
 
\begin{document}

 \title{Plane wave holonomies in loop quantum gravity II: sine wave solution}

\author{Donald E. Neville}

\email{dneville@temple.edu}

\affiliation{Department of Physics,
Temple University,
Philadelphia 19122, Pa.}


\newcommand{\rA}{\mathrm{A}}
\newcommand{\rC}{\mathrm{C}}
\newcommand{\rD}{\mathrm{D}}
\newcommand{\rE}{\mathrm{E}}
\newcommand{\rF}{\mathrm{F}}
\newcommand{\rH}{\mathrm{H}}
\newcommand{\rK}{\mathrm{K}}
\newcommand{\rL}{\mathrm{L}}
\newcommand{\rM}{\mathrm{M}}
\newcommand{\rN}{\mathrm{N}}
\newcommand{\rP}{\mathrm{P}}
\newcommand{\rR}{\mathrm{R}}
\newcommand{\rS}{\mathrm{S}}
\newcommand{\rU}{\mathrm{U}}
\newcommand{\rV}{\mathrm{V}}
\newcommand{\rY}{\mathrm{Y}}
\newcommand{\rmd}{\mathrm{d}}
\newcommand{\rmi}{\mathrm{i}}

\newcommand{\E}[2]{\mbox{$\rE^{#1}_{#2}$}}
\newcommand{\Eav}{\mbox{$\bar{\rE}^z_Z$}}
\newcommand{\A}[2]{\mbox{${\rA}^{#1}_{#2}$}}
\newcommand{\Nt}{\mbox{\underline{N} }  }
\newcommand{\Ht}{\mbox{$\tilde{\mathrm{H}} $} }
\newcommand{\Etwo}{\mbox{$^{(2)}\tilde{\mathrm{E}} $}\ }
\newcommand{\Etld }{\mbox{$\tilde{\mathrm{E}}  $}\ }
\newcommand{\Vtwosq}{\mbox{$(^{(2)}{\mathrm{V}})^2 $}}
\newcommand{\Vtwo}{\mbox{$^{(2)}{\mathrm{V}} $}}
\newcommand{\Ethree}{\mbox{$^{(3)}{\mathrm{E}} $}}
\newcommand{\etwo}{\mbox{$^{(2)}{\mathrm{e}}\, $}}
\newcommand{\ethree}{\mbox{$^{(3)}{\mathrm{e}} $}}


\newcommand{\nn}{\nonumber \\}
\newcommand{\rta}{\mbox{$\rightarrow$}}
\newcommand{\rla}{\mbox{$\leftrightarrow$}}
\newcommand{\eq}[1]{equation (\ref{#1})}
\newcommand{\Eq}[1]{Equation (\ref{#1})}
\newcommand{\eqs}[2]{equations (\ref{#1}) and (\ref{#2})}
\newcommand{\Eqs}[2]{Equations (\ref{#1}) and (\ref{#2})}
\newcommand{\Or}{\mbox{O }}
\newcommand{\bra}[1]{\langle #1 \mid}
\newcommand{\ket}[1]{\mid #1 \rangle}
\newcommand{\braket}[2]{\langle #1 \mid #2 \rangle}
\newcommand{\cd}{\delta_{(c)}\,}
\newcommand{\td}{\tilde{\delta}\,}
\newcommand{\si}{\mbox{sgn}}

\begin{abstract}

    This paper
constructs an approximate sinusoidal wave packet
solution to the equations of loop quantum gravity (LQG).
There is an SU(2) holonomy on each edge of the
LQG simplex, and the goal is to study the behavior
of these holonomies under the influence of a passing
gravitational wave.
The equations are solved in a small sine approximation:
holonomies are expanded in powers of sines and
terms beyond $\sin^2$ are dropped; also, fields vary slowly
from vertex to vertex. The wave is
unidirectional and linearly polarized. The Hilbert space
is spanned by a set of
coherent states tailored to the symmetry of the
plane wave case.  Fixing the spatial diffeomorphisms
is equivalent to fixing the spatial interval between
vertices of the loop quantum gravity lattice.  This
spacing can be chosen
such that the eigenvalues of the triad operators
are large, as required in the small sine limit,
even though the holonomies are  not large.
Appendices compute the energy of the wave, estimate the
lifetime of the coherent state packet, discuss
coarse-graining, and determine the behavior of the 
spinors used in the U(N) SHO realization of LQG.

\end{abstract}


\maketitle

\section{Introduction}

    This is the second of two papers with the goal
of developing some intuition for the behavior of
LQG holonomies and fluxes in the presence of a gravitational wave.
The previous paper (paper I) constructs a
classical LQG theory having planar symmetry
\cite{Semiclassical}.  The gravitational excitation
is assumed to be unidirectional
and singly polarized. Constraints are
evaluated in a small sine (SS), slow variation (SV) limit.
Holonomies are expanded in powers of sine,
\begin{eqnarray*}
    \mathbf{h}(\theta, \hat{n})^{1/2} &=& \cos(\theta/2)\mathbf{1} + i\sin(\theta/2)\hat{n}\cdot \mathbf{\sigma}\\
    &\cong& \mathbf{1} + i\sin(\theta/2)\hat{n}\cdot \mathbf{\sigma} + \Or \sin^2,
\end{eqnarray*}
where h is a spin 1/2 holonomy, a rotation around axis $\hat{n}$
through angle $\theta$.  Terms of order $\sin^3$ and higher
in the constraints are dropped. Dynamical
functions f are assumed to vary slowly from
vertex to vertex: $\delta f/f \ll 1$. The two
assumptions, small sine and slow variation,
are closely connected, and for brevity sometimes we will
refer to them simply as the small sine ((SS) approximation.

    Paper I imposed all gauges at the
classical level, except the spatial diffeomorphism gauge.
A diffeomorphism
gauge is chosen in section \ref{DiffGauge} of the present
paper.  Some discussion is
required; the value of the gauge fixing constant C
is closely connected to peak angular
momentum of the coherent states.

    Section \ref{ScalarConstraint}
quantizes the theory.  As emphasized in this
section (and in paper I), any classical solution to the
constraints yields a corresponding solution
to the quantum constraints, since we use
coherent states as a basis for the Hilbert
space.  Section \ref{Undamped}
constructs such an approximate classical solution,
an undamped sine wave.
Section \ref{Damping} adds the damping.

        Section \ref{Coherent} sketches the construction
of the coherent states.
Section \ref{FreeParticle},
compares the SU(2) coherent states to the familiar
coherent states for the free particle.  This analogy
is used to justify the form of
the SU(2) states, in a manner which is
qualitative, but should be intuitively convincing.
Full details of
the construction  are given in reference \cite{1};
see also \cite{Vol1}.
Section \ref{MatrixElements}
summarizes the most important matrix elements of the SU(2)
coherent states.

    The coherent states depend upon a number of
angle and angular momentum parameters. Section
\ref{FixParameters} determines parameter values
such that the expectation values of the triads
reproduce the sinusoidal solution constructed in
section \ref{Damping}.

    Appendix \ref{ADM} computes the ADM energy of the wave.
Appendix \ref{StrongSpreading} estimates the lifetime
of coherent state wave packets.  In LQG the lifetime of the
packet depends on the standard deviation  of the
angular momenta at each vertex,  rather than the
usual, standard deviation of velocities in the
packet.  Appendix \ref{Renormalization} discusses
coarse-graining.  The present solution is an especially
simple example, illustrating the method of coarse-graining
proposed in references \cite{Friedel} - \cite{Livine}.

\subsection{Plane waves in classical general relativity}
\label{ClassicalGauges}

    The classical literature uses primarily two
gauges: the one used in this paper, in which $ -g_{tt} = g_{zz} = 1$,
$g_{\mu\nu} = g_{\mu\nu}(t,z)$; and a gauge
\[
    ds^2 = dx^2 + dy^2 + dz^2 - dt^2 + f(u,x,y)du^2,
\]
u = (z - t)/$\sqrt{2}$.  The first gauge was used
by Baldwin and Jefferys in their pioneering paper \cite{Brinkmann}.
Peres derived an exact solution for an undamped
sinusoidal plane wave
using the second gauge \cite{Peres}.
The Peres solution, when converted to the gauge
used in this paper, becomes the undamped solution
of section \ref{Undamped}.   Griffiths \cite{Griffiths} shows
how to convert between the two gauges and describes
additional exact non-sinusoidal solutions.

\section{Fixing the Diffeomorphism Gauge}
\label{DiffGauge}

    There is an apparent contradiction between
two assumptions made in reaching the small sine
limit.  Coherent states work best when eigenvalues
are large; yet fields must be weak.  How can fields be
small, if eigenvalues are large?

    The LQG \Etld operators contain area factors
not present in their field theory (FT) analogs:
\begin{equation}
    (2/\kappa\gamma)\,\mathrm{E}^i_I (\mbox{FT)}\, \rta (2/\kappa\gamma)\,\Delta x^j \wedge \Delta x^k \,\mathrm{E}^i_I \;(\mbox{LQG)}.
\label{QuantumE}
\end{equation}
In this paper "Field theory" refers to a classical theory
based on fluxes and connections which have support on
the continuum; LQG refers to a classical or quantum
theory based on fluxes and holonomies which have support on
a lattice.)  The area is in Planck lengths squared, because of the
$\kappa\gamma$ factor.
Suppose the $\Delta x^i$ are taken to be $10^2$
Planck lengths (an extremely tiny length,
by classical standards).  The classical
triad  may be order unity; yet the quantum
eigenvalue will be order $10^4$.  Therefore, \emph{
typical angular momenta in the wavefunction can
be order $10^4$}, far from order unity,
\emph{even though classical values are order unity}.
This fact resolves the apparent contradiction
discussed in the previous paragraph.

As in the literature for classical general relativity, the
diffeomorphism gauge is chosen such that
$\mathrm{g}_{zz} = 1$.  In the notation of
paper I, this gauge has parameter p = 1/2.
\begin{eqnarray}
    \Etwo(FT) &=& (\mathrm{C}_{FT} \,\mathrm{E}^z_Z)^{p + 1/2} \nn
        &=& \mathrm{C}_{FT}\, \mathrm{E}^z_Z ; \:\mbox{equivalently,} \nn
     (e^Z_z)^2 &=& \si(e) \,\mathrm{C}_{FT}.
\label{ClGaugeChoice}
\end{eqnarray}
$\mathrm{C}_{FT}$ is a constant.  \Etwo is the determinant of the 2 x 2 transverse
(x,y) triads. The last line of \eq{ClGaugeChoice} expands the \Etld
in triads.   sgn(e) is the sign of the  3 x 3 triad
determinant.
\[      e = \si(e) \,|e|.        \]
Since $e^Z_z$ must match to flat
space at the front of the packet,
\begin{eqnarray}
    \mathrm{C}_{FT} &=&  \si(e) ;\nn
    e^Z_z &=&  \pm 1 := \si(z)
\label{CclChoice}
\end{eqnarray}

    Now take this over to loop quantum gravity.  The spatial diffeomorphism gauge
must be chosen such that, when factors of $\Delta x^i$
are stripped out, one recovers the classical gauge fixing.
\begin{eqnarray}
    \Etwo(LQG) &=& \mathrm{C}_{LQG} \,\mathrm{E}^z_Z(LQG); \nn
        \mathrm{C}_{LQG}&=&  (\Delta z)^2 \,\mathrm{C}_{FT}  \nn
\label{QuGaugeChoice}
\end{eqnarray}
Each $\mathrm{E}^x$ in \Etwo(LQG) will have an area
factor $\Delta y \Delta z$; each $\mathrm{E}^y$ in \Etwo will
have a factor $\Delta x \Delta z$.  The factors of $\Delta x, \Delta y$
are also present in $\mathrm{E}^z$, but not the $\Delta z$.
Therefore the missing $\Delta z$ factors turn up in
$\mathrm{C}_{LQG}$.  $\mathrm{C}_{FT}$  is still the classical
value, sgn(e).  When we pick $\mathrm{C}_{LQG}$, we are picking
a  value for $\Delta z$ (in Planck units, after both sides of \eq{QuGaugeChoice}
are divided by $(\kappa\gamma/2)^2$; compare  \eq{QuantumE}.)

    Caps, $\Delta X^I$, denote  Local Lorentz
coordinates; lower case $\Delta x^i$ denote coordinates on
the global manifold.  The two sets of coordinates are related.
\begin{eqnarray}
    \tilde{\rE}^i_I(LQG) &=& \si(e) \, (e  e^i_I)(FT) \Delta x^j \,\Delta x^k \nn
                    & =& \si(e) \,e^J_j e^K_k \,\Delta x^j \,\Delta x^k \nn
                    & =& \si(e) \, \Delta X^J \,\Delta X^K.
\label{areaEqArea}
\end{eqnarray}
\Eq{areaEqArea}  may also be written as
\begin{equation}
    \Etld^i(FT)\Delta x^j \Delta x^k
                = \si(e) \, \Delta X^J \,\Delta X^K.
\label{areaEqArea2}
\end{equation}

    The weak-classical-field-but-large-eigenvalue connection
emerges if one multiplies the last equation by $(2/\kappa\gamma)$
and equates the result to a spin eigenvalue j.
\[
    (2/\kappa\gamma)\Etld^i(FT)\Delta x^j \Delta x^k = \Or j.
\]
j can be large, even though \Etld(FT) is small, because of the large area.

    This way of introducing j eigenvalues can be thought of as the reverse of
the familiar discussion.  Usually, one imagines a given spin
network having given spin j and applies \Etld.  The end result
is an area.  Here that process is reversed.  One adjusts the area
until the end result is the desired (large) spin value j.

   The  \Etld(FT) in \eq{areaEqArea2} will have z dependence, which means that the
$\Delta x^j \Delta x^k$ vary with z, or the $\Delta X^J \,\Delta X^K$,
or both.  We assume the global coordinates $x^i$ are fixed; the
variation is in the Lorentz lengths $X^I$.  Equivalently,
\Etld(FT) and \Etld(LQG) have the
same variation with z, since
\mbox{\Etld(FT)} and \Etld(LQG) differ only by factors of $\Delta x^j$,
which are held fixed.

    Support for this assumption comes from a later result in the sections on coherent
state parameters. The coherent states are approximate eigenstates of
the \Etld(LQG) in \eq{areaEqArea}, with eigenvalues
equal to an angular momentum or Z coordinate
of angular momentum.
\[
    \tilde{\rE}^a_A(LQG)\ket{\mbox{coh}} = (\kappa\gamma/2) \, \rL^a_A \,\ket{\mbox{coh}};,\quad a = x,y;
\]
\begin{equation}
    \tilde{\rE}^z_Z (LQG)\ket{\mbox{coh}} = (\kappa\gamma/2)\, m \,\ket{\mbox{coh}}.
\label{EeqAngularMomenta}
\end{equation}   The last line of \eq{areaEqArea} gives
\begin{equation}
    (\kappa\gamma/2)(\rL^a_A \: \mbox{or} \: m) = \si(e) \, \Delta X^J \,\Delta X^K.
\label{LeqAreas}
\end{equation}
If the Lorentz lengths $\Delta X^I$
are taken as fixed, then the canonical momenta cannot
vary in the presence of a gravitational wave, a reduction
to the absurd.

     When quantizing plane waves in FT, using ADM
variables, one can renormalize constraints by
dividing out a factor $\Delta x \Delta y$.  The FT
expressions then contain only integrals over z.
Such a renormalization is
not possible in LQG,
because not every term contains an overall factor
of $\Delta x \,\Delta y$.  Some $(\Delta x, \Delta y)$
are hidden in holonomies and do not cancel out.
In FT the integrals over transverse
directions are infinite, and  renormalization
is mandatory.  In LQG the transverse integrals
range over the
circumferences of the x and y circles and are finite.
Renormalization is not necessary.

\subsection{N Can be Fixed at Unity}

    Paper I introduced a modified lapse \Nt which
obeys simpler boundary conditions than the usual lapse N.
However, it is desirable to
arrange N $>$ 0, so that dt and dT "run"
in the same direction.
\begin{eqnarray}
    \mathrm{dT} &=& e^T_t \rmd t = \mathrm{N} \, \rmd t; \nn
    e^T_i &=& 0; \:i = x,y,z.
\label{dtEqdT}
\end{eqnarray}
The second line is the usual gauge choice
which reduces full Lorentz symmetry to SU(2).

      Fortunately, the diffeomorphism
gauge chosen above leads to a simple
relation between \Nt and N.
\begin{eqnarray}
    \Nt(\mathrm{FT}) &:=& \mathrm{N} (\mathrm{E}^z_Z/\mid e \mid)_{\mathrm{FT}} \nn
        &=& \mathrm{N}\,e^z_Z = \mathrm{N}\, \si(z).
\label{FixN}
\end{eqnarray}
sgn(z) is the sign of $e^Z_z$ and \E{z}{Z}.  Also, the
constraints of paper I require
\Nt to be a constant: $\cd \Nt = 0$.
If the constant is chosen appropriately, N becomes unity:
\begin{equation}
    \Nt(\mathrm{FT}) =  \si(z).
\label{FixNt}
\end{equation}
A corollary: with this choice, neither \Nt nor N can vanish. $\Box$

    The LQG \Nt is constructed from the definition \eq{FixNt}
except use LQG \Etld.
\begin{equation}
    \Nt(\mathrm{LQG}) = \Nt(\mathrm{FT})/\Delta z = 1/\Delta Z .
\label{QuNt}
\end{equation}
\Nt is a contravariant rank one tensor, therefore
needs a $1/\Delta z$ to make it diffeomorphism invariant.

\subsection{\Nt May Be Chosen Unity}

    It is convenient to make the light cone variable du equal to
the inertial frame dU,
\begin{equation}
    \rmd u = (e^z_Z \,\rmd Z - e^t_T\, \rmd T)/\sqrt{2} = (\si(z)\, \rmd Z - \rmd T)/\sqrt{2},
\label{duEqdU}
\end{equation}
This requires
\begin{equation}
    \si(z) = + 1 = \Nt(\mathrm{FT}) .
\label{FinalGaugeChoice}
\end{equation}

\subsection{Some Triads Can Vanish}

    The gauge choice \eq{CclChoice} forbids zeros
of $e^Z_z$; but not zeros of \etwo, the determinant
of the transverse $e^A_a$.  \Etwo and \E{z}{Z}
each contain one power of \etwo and could conceivably
pass through zero simultaneously, when away from
the small sine limit.

\section{The quantum scalar constraint}
\label{ScalarConstraint}

    Our final formula for the scalar constraint \Ht in
paper I was
\begin{eqnarray}
    \Ht   & = &\sum_n (1/\kappa)\{(1/2)(\cd \mathrm{E}^y_Y /\mathrm{E}^y_Y   -\cd \mathrm{E}^x_X /\mathrm{E}^x_X)^2 \mathrm{E}^z_Z\\
              &&\quad +\cd \mathrm{E}^z_Z \,[- \,(\cd \Etwo)/\Etwo \nn
        &&\quad + \,\cd \mathrm{E}^z_Z/2\mathrm{E}^z_Z] + \cd(\cd \mathrm{E}^z_Z) \} \\
            &=&  0.
\label{H3}
\end{eqnarray}
The gauge choice \eq{QuGaugeChoice} implies
\begin{eqnarray}
    \cd \Etwo/\Etwo &=& \cd \mathrm{E}^z_Z/\mathrm{E}^z_Z \nn
                &=& \cd \mathrm{E}^x_X /\mathrm{E}^x_X + \cd \mathrm{E}^y_Y /\mathrm{E}^y_Y.
\label{EliminateEzZ}
\end{eqnarray}
This and the next equation use a distributive law for the
difference which is valid given the slow variation (SV) assumption.
\[ \cd(AB) = (\cd A) B + A\, (\cd B).   \quad \mbox{(SV)}  \]
In the constraint, \eq{H3}, One can divide through by $\mathrm{E}^z_Z$ and use
\eq{EliminateEzZ} to eliminate $\cd \rE^z_Z$.  The double difference
may be rewritten using
\begin{eqnarray}
     \cd\{\cd \mathrm{E}^z_Z\}/\mathrm{E}^z_Z &=& \,\cd \{[\cd \mathrm{E}^x_X /\mathrm{E}^x_X + \cd \mathrm{E}^y_Y /\mathrm{E}^y_Y] \mathrm{E}^z_Z \}/\mathrm{E}^z_Z \nn
        &=& \,[\cd (\cd \mathrm{E}^x_X) /\mathrm{E}^x_X + \cd (\cd \mathrm{E}^y_Y) /\mathrm{E}^y_Y] \nn
        && - [\cd \mathrm{E}^x_X /\mathrm{E}^x_X]^2 - [\cd \mathrm{E}^y_Y /\mathrm{E}^y_Y]^2 \nn
        && + [\cd \mathrm{E}^x_X /\mathrm{E}^x_X + \cd \mathrm{E}^y_Y /\mathrm{E}^y_Y]^2  .
\label{ExpandDoublecd}
\end{eqnarray}
The constraint simplifies to
\begin{equation}
    0 = \cd (\cd \mathrm{E}^x_X)(1 /\mathrm{E}^x_X) + \cd (\cd \mathrm{E}^y_Y)(1 /\mathrm{E}^y_Y).
\label{H4}
\end{equation}
At this point one can make the transition from classical to quantum.
Classical functions are replaced by quantum operators; brackets become
quantum commutators (Dirac rather than Poisson brackets, because
the unidirectional constraints are second class).  There is
a factor ordering question, because Dirac
brackets imply the \Etld no longer commute with themselves.  In a
typical LQG quantization involving Poisson brackets, the \Etld are
ordered to the right of the K's.  Here,
the $\cd$\Etld are equivalent to K's because of
the unidirectional constraints.  Therefore triads have
been moved to the right of the $\cd$\Etld.

\subsection{Comparison to Classical Results}

    The classical calculation yields the following
results for the non-zero components of the Einstein
and Weyl tensors.
\begin{eqnarray}
   \mathrm{G}_{uu}&=& \ddot{\mathrm{E}}^x_X /\mathrm{E}^x_X + \ddot{\mathrm{E}}^y_Y /\mathrm{E}^y_Y = 0; \nn
   \mathrm{C}^x_{uxu} &=&\ddot{\mathrm{E}}^x_X /\mathrm{E}^x_X - \ddot{\mathrm{E}}^y_Y /\mathrm{E}^y_Y = -\mathrm{C}^y_{uyu}.
\label{WeylEinsteinEDoubleDot}
\end{eqnarray}
Variables are x,y,u,v.  Fields are single polarization and unidirectional
(dependent  on u only); dots denote derivatives with respect to u.
Gauge is $e^Z_z = \pm 1$.  The LQG constraint \eq{H4} is just the
classical constraint, with u derivatives replaced by z differences.

    From \eq{WeylEinsteinEDoubleDot}, the classical Weyl tensor is the
scalar constraint, with one minus sign change.  This same
relation (between scalar constraint and Weyl) holds
in the quantum case.  Therefore
\begin{eqnarray}
    \cd (\cd \mathrm{E}^x_X) /\mathrm{E}^x_X  &=& -\cd (\cd \mathrm{E}^y_Y) /\mathrm{E}^y_Y; \nn
    \mbox{Weyl} &=& 2 \cd (\cd \mathrm{E}^x_X) /\mathrm{E}^x_X .
\label{WeylFromEinstein}
\end{eqnarray}
One can pick a desired curvature, choose an $\mathrm{E}^x_X$ which
produces this curvature, and immediately
have a solution to the scalar constraint.

\subsection{Coherent states, Dirac brackets, and the Scalar Constraint}

    In leading order, coherent states do not preserve quantum commutators.
Let $O_1, O_2$ be two quantum operators peaked at values
$O_i(cl)$.  Then
\begin{eqnarray}
    \bra{\mbox{coh}}O_1\, O_2\ket{\mbox{coh}} &=& \bra{\mbox{coh}}O_1 \ket{\mbox{coh}}\bra{\mbox{coh}} O_2\ket{\mbox{coh}} \nn
    &&\quad + \sum_{SC} \bra{\mbox{coh}}O_1 \ket{\mbox{SC}}\bra{\mbox{SC}} O_2\ket{\mbox{coh}} \nn
    &\cong& O_1(cl)  O_2(cl),
\label{Classical=Quantum}
\end{eqnarray}
The $O_i$ acting on a coherent state typically give
back the coherent state, plus small correction (SC) states
which are down by order $1/\sqrt{\rL} $ \cite{1}.
If we neglect the SC states, then
the commutator $< [O_1, O_2]>$ is zero.
Thiemann and Winkler, without constructing the SC states,
have shown that Poisson brackets
are preserved in the semiclassical limit \cite{GCSIII}
in the sense that the quantum commutator is given by i$\hbar$ times
the classical Poisson bracket.
Since Dirac brackets are functions of Poisson brackets,
it is likely that Dirac brackets are preserved also.

    The tendency of coherent states to turn quantum
operators into classical expressions is helpful in another
context.  If one has a classical solution to the scalar
constraint, one immediately has a quantum solution.
\[
    (H = O_1\, O_2 \cdots)\ket{\mbox{coh}} \cong (O_1(cl)  O_2(cl) \cdots)\ket{\mbox{coh}}.
\]
The next two sections construct such a classical solution.

\section{A gravitational sine wave}
\label{Undamped}

    The next two sections construct a sine wave solution.
The solution is classical, but (as just mentioned) can
be translated immediately into a solution to the quantum constraint.
The  first step (this section) constructs a
solution which is periodic,
but undamped.  The following section adds damping.

    The undamped solution is
\begin{eqnarray}
    \mathrm{E}^x_X(LQG;n)&=& (\Delta z \Delta y) \si(x)\{1 - a \sin \,[(2\pi n/\rN _\lambda) \,]/2!  \nn
                & & -(a^2/32)[ \,\cos(4\pi n/ \rN _\lambda) + (4\pi/\rN _\lambda)^2(n)^2/2 \,]\}.
\label{QuSineWave}
\end{eqnarray}
a is a small, dimensionless, constant amplitude. $\rN _\lambda$ is a constant,
the number of vertices in a length equal to one wavelength.
When n changes by $\rN _\lambda$, the phase of
the sine changes by 2$\pi$. sgn(x) = $\pm 1$
is the sign of $e^x_X$, -1 if the x and X axes increase in opposite directions.
The order $a^2$ terms are freqency-doubled, a typical non-linear effect.

    From the expression for the Einstein tensor G, the
expression for \E{y}{Y} must have a linear in a term
identical to \eq{QuSineWave}, except a $\rta$ -a
(and $x \leftrightarrow y$) .

    With a slight abuse of a standard notation, one can define a
k vector in n space, i.e. a vector which gives
the change in  phase per unit change in n.
\begin{eqnarray}
    (2\pi /\rN _\lambda)&:=&  k;\nn
    k\, n &=& (k/\mid \Delta Z \mid)(n \mid \Delta Z \mid) = (2 \pi /\mbox{wavelength}) (\mid Z \mid)
\label{defk}
\end{eqnarray}
The second line gives the connection to the usual k, the
change in phase per unit change in length.

The above solution is approximate because an exact solution
requires an infinite series, whereas the quantum
solution of \eq{QuSineWave} stops at order $a^2$.

    To check the constraint and compute curvature, one must
compute $\delta^{(2)}E/E$.  The second difference of the linear-
in-a term, \eq{QuSineWave}, is
\begin{align}
   -(a/2) \si(x)&\{\sum_{\pm}\sin[(2\pi/\rN _\lambda)(n\pm1)] - 2 \sin(2n\pi/\rN _\lambda)\}(\Delta z \Delta y) \nn
    &= -(a/2)\si(x)\{ \sin(2n\pi/\rN _\lambda)(2\cos(2\pi/\rN _\lambda) - 2)\}(\Delta z \Delta y).
\label{LinearIna}
\end{align}
The first sine was expanded using
$\sin(A \pm B) = \sin A \cos B \pm \cos A \sin B$.
To estimate the size of $\rN _\lambda$, one can use the connection between
$\rN _\lambda$ and the classical wavelength, \eq{defk}.  Since that
wavelength is macroscopic, whereas $\Delta Z$, the change in z per
unit change in n, is of order a few hundred Planck lengths,
$\rN _\lambda$ must be astronomically large,  and 1/$\rN _\lambda$ must be negligible, except
when multiplied by n.  Therefore one can expand the cosine
in \eq{LinearIna}, and the second difference becomes
\begin{equation}
    (a/2)\si(x)(2\pi/\rN _\lambda)^2 \sin(2 \pi n/\rN _\lambda) (\Delta z \Delta y).
\label{LinearIna2}
\end{equation}

    The term quadratic in a, \eq{QuSineWave}, is handled
similarly: trigonometric identities are used to expand functions of
n $\pm$ 1; functions of 1/$\rN _\lambda$ are power-series expanded.
The total second difference (both linear and quadratic in a) is
\begin{eqnarray}
    \delta^{(2)}\rE^x_X(LQG;n)& =& (a/2)\si(x)(2\pi/\rN _\lambda)^2 \sin(2n\pi/\rN _\lambda) \nn
          &&  \times [1 - (a/2) \sin(2n\pi/\rN _\lambda)](\Delta z \Delta y).
\label{SecondDifference}
\end{eqnarray}
The square bracket is \E{x}{X}(FT), so that the curvature is
\begin{equation}
    2 \delta^{(2)}\rE^x_X(LQG;n)/\rE^x_X =2 (a/2)(k)^2 \sin(k n),
\label{QuCurvature}
\end{equation}
where we have shifted to the new k vector $ 2 \pi/\rN _\lambda$.
The calculation for
\E{y}{Y} is identical, except (a $\rta$ -a); the term linear in a
changes sign, but not the term quadratic in a.   Then from \eq{SecondDifference}
with a $\rta -a$,  the scalar constraint \eq{H4} is satisfied.

    The role of the small amplitude a needs to be clarified.
In deriving \eq{QuCurvature} one may assume that the curvature
is \emph{linear} in amplitude a, while transverse \Etld are
infinite series in a.  I.~e., order $a^2$ and higher corrections to
curvature vanish.  To see how this happens,
rewrite the expression for the \Etld in a manner which
emphasizes the dependence on a.
\begin{eqnarray}
    \delta^{(2)}\rE^x/\rE^x &=& ( \ddot{B}_1 + \ddot{B}_2 + \cdots) /(1 + B_1 + \cdots) \nn
                        &=&  \ddot{B}_1 + \ddot{B}_2 - B_1 \ddot{B}_1 + \cdots.
\label{NoSecondOrder}
\end{eqnarray}
$ B_p$ is order $a^p$; double dots indicate
second differences; and $\cdots$
indicate terms which contribute  cubic and higher
terms to the curvature.  Choose $B_2$ such that
\[
    ( \ddot{B}_1 + \ddot{B}_2 + \cdots) /(1 + B_1 + \cdots) = \ddot{B}_1 \,(1 + B_1 + \cdots)/(1 + B_1 + \cdots),
\]
Equivalently, choose $B_2$ so that the order $a^2$ terms in
\eq{NoSecondOrder} cancel.
\begin{equation}
    \ddot{B}_2/B_1 = \ddot{B}_1.
\label{LowestOrder}
\end{equation}
Then the order $a^2$ contributions to curvature vanish.

    Most results will be calculated only to order a.  Note
one exception, however. For
the above mechanism to work, the \Etld must be known
to order $a^2$.

    One can generalize \eq{LowestOrder} to cubic and higher orders
in a.  Given $B_1, B_2,\cdots,B_{p-1}$, determine $B_p$ by
solving the equation
\begin{equation}
    \ddot{B}_p/B_{p-1} = \ddot{B}_1.
\label{HigherOrder}
\end{equation}
Then
\begin{eqnarray}
     ( \ddot{B}_1 + \cdots +\ddot{B}_p) &/&(1 + B_1 + \cdots + B_{p-1}) \nn
      &=& \ddot{B}_1 \,(1 + \cdots + B_{p-1})/(1  + \cdots + B_{p-1})  \nn
     &=& \ddot{B_1}.
\label{HighOrderCancelation}
\end{eqnarray}
The curvature is order a, to all orders.

    Let \eq{HighOrderCancelation} represent  the series for
$\ddot{\rE^x_X}/\rE^x_X$.   There is another one for
$\ddot{\rE^y_Y}/\rE^y_Y$ with $B_1 \rta -B_1$, in order for
the Einstein tensor to vanish in order a.  From the
recurrence relation \eq{HigherOrder}, in the y series \emph{all}
terms with odd powers of a have the opposite sign.

    To make contact with the classical curvature, \eq{WeylEinsteinEDoubleDot},
divide the second difference by ($\Delta \mathrm{U})^2$, then
convert differences to derivatives with respect to
U.  From $U = (Z - T)/\sqrt{2}$,
\[
    \Delta \mathrm{U} = \Delta \mathrm{Z}/\sqrt{2},
\]
in a formalism where T is held constant.  Then
\begin{eqnarray}
    \mathrm{C}^x_{uxu}(cl) &=& \{\delta^{(2)}\rE^x_X(LQG;n)/[\rE^x_X(LQG;n) ] -(x \rta y)\}/(\Delta \rU)^2 \nn
        &=& 2 (a/2)(k)^2 \sin(k n)/(\Delta U)^2 \nn
        &=& a (2 \pi \Delta Z/\lambda)^2 \sin(k n) 2/(\Delta Z)^2 \nn
        &=& 2a (2 \pi/\lambda)^2 \sin(2 \pi z/\lambda),
\label{QutoClassicalWeyl}
\end{eqnarray}
where $\lambda$ is the classical wavelength.

    In this section \E{a}{A}(cl.) was chosen
to start off with leading term +1.   This choice, together
with the gauge choice $e^Z_z = +1$, implies
sgn(e) = +1.  To obtain the opposite choice, sgn(e) = -1,
change one \E{b}{B} to - \E{b}{B}.  The
new solution leaves the Weyl tensor unchanged
and continues to satisfy the \Ht = 0 constraint.

\subsection{A second solution}

    Since we are dealing with second order difference
equations, there should be a second solution, in addition
to the solution given at \eq{QuSineWave}.  In the theory
of second order differential equations, the two series
solutions around z = 0 have leading powers 1 and z.
The first series fits the function at z = 0; the
second series fits the first derivative.

    By analogy, one would expect two solutions to the difference
equation, with leading powers $B_0$ =
1 and $B_0$ = (2 $\pi$ n/4 q) $:=$ k n.  With this hint, plus

\[
    \delta^{(2)}[B_0 + B_1]/[B_0 + \cdots] =  a\sin(k n),
\]
one can construct a second solution. It has $B_0$ and $B_1$ terms
\begin{equation}
    k n - a [(k n)\sin(k n) +2 \cos(k n)]/k^3 + \cdots.
\label{SecondSolution}
\end{equation}
This solution would be needed if the  difference of $\rE^a$
were non-zero at infinity.  Since the difference vanishes,
this solution can be ignored.

\section{Inclusion of damping}
\label{Damping}

    The solution \eq{QutoClassicalWeyl} is
infinite in length.  The solution may be made into a packet by
including damping factors.
\begin{eqnarray}
    \mathrm{E}^x_X(LQG;n)&=& (\Delta z \Delta y)\si(x) \{1 - (a/2) \,\exp( \mp \rho \, n ) \,\sin[k\,  n\mp \phi]  \nn
                & & + \,(-a^2/32)[ \,\exp( \mp 2\rho \,n ) \,\cos(2 k \,n \mp 2 \phi) \nn
                &&+ \,( \,\exp \,[\mp 2\rho \, n ] \pm 2 \rho \,n  -1) \,(f^2/\rho^2)  \,\cos \phi]\};
\label{ExponentialPacket}
\end{eqnarray}
\[
    f^2 := (k^2+\rho^2).
\]
Upper (lower) sign refers to n $>$ 0 (n $<$ 0).
For simplicity in what follows, The discussion to follow  will consider only
the case n $>$ 0 (upper sign); the n $<$ 0 follows by changing
\begin{equation}
    \rho \rta - \rho; \: \phi \rta -\phi.
\label{nVsMinusn}
\end{equation}
The expression for $\mathrm{E}^y_Y(LQG;n)$ is \eq{ExponentialPacket}
with x $\rla$ y and a $\rta$ -a.

    The exponential damping factors have discontinuities in derivative at
n = 0; and from \eq{nVsMinusn} the angle $\phi$ is
undefined at n = 0.  A discontinuity by itself is not a problem
because the damping function is defined only  at discrete
points.  The problems at n  = 0 turn out to be
minor; the  value n = 0  is treated in section \ref{n=0}.

    If the curvature is to remain a sine wave with
zero phase, then \Etld must include a constant phase $\phi$.  When one solves
the differential equation F = ma for the damped oscillator,
one finds that each derivative shifts the phase by more than
the usual $\pi/2$.
\begin{align}
   (\rmd/\rmd t)[\exp(-\rho t)\sin(\omega t - \phi)] &= \sqrt{\omega^2 + \rho^2}\exp(-\rho t)\cos(\omega t - \phi+\psi);\nn
    \cos \psi &= \omega/\sqrt{\omega^2 + \rho^2}.
\end{align}
Exactly the
same phenomenon occurs in the difference case.  One may choose
a non-zero phase $\phi $ for \Etld, $\phi$ to be determined.
The differences shift this phase, until
the curvature becomes a sine wave with
zero phase.

    Computation of the damped second difference is straightforward.
As before, sinusoidal functions of n $\pm $1
are expanded using trigonometric identities.  As before, k is assumed
small and functions $\sin k$, $\cos k$
are power series expanded.  A new feature: the damping
parameter $\rho$ is assumed small compared to wavelength,
\[
    \rho/k \ll 1,
\]
so that
functions $\exp(-\rho)$ may be power-series expanded, whenever
$\rho$ is not multiplied by n.  Since 1/$\rho$ measures the
length of the packet, small $\rho$/k implies the packet
contains many wavelengths.

    The second
difference of the term linear in a is
\begin{align}
    (a/2) \exp(- \rho \, n ) &\{ \sin (k\, n - \phi) [k^2 - \rho^2] \nn
     &+ 2 k \,\rho \cos(k\, n - \phi)\}(\Delta Z)^2[1 + \mbox{order}\: k^2, \rho^2, k \rho\,]; \:n \neq 0.
\end{align}
Now choose $\phi$ so that the linear-in-a term (and
ultimately, the curvature) collapses to a $\sin (k\, n)$ times a
damping factor.
\begin{eqnarray}
    \cos \phi &=& (k^2 - \rho^2)/f^2; \nn
    \sin \phi &=& 2 \,k \,\rho/f^2;\nn
    f^2 &=& k^2 + \rho^2.
\label{defPhi}
\end{eqnarray}
The second difference of the linear-in-a term reduces to
\[
    (a/2) \exp(- \rho \, n ) \,f^2 \,\sin (k\, n) \,(\Delta Z)^2.
\]

    The term quadratic in a, \eq{ExponentialPacket},
requires one extra trigonometric identity.  After the usual
expansions, that second difference becomes
\begin{align}
   -(a^2/8)\exp(- 2\rho \, n )&\{-(k^2 - \rho^2)\cos(2 k \,n - 2 \phi)\nn
           &+ (2\rho\, k) \sin(2k \,n - 2 \phi) + f^2 \cos \phi\}(\Delta Z)^2\nn
            &= -(a^2/8)f^2\exp(- 2\rho \, n )\{-\cos (2 n \,k - \phi)+ \cos \phi\}(\Delta Z)^2\nn
            &= -(a^2/4)f^2\exp(- 2\rho \, n )\{\sin(k \,n - \phi) \sin(k\, n)\}(\Delta Z)^2.
\end{align}
The last line uses the identity
\[
    2 \sin A \sin B = \cos (A-B) - \cos (A+B).
\]
One can now factor out
\[
    \mathrm{E}^x_X = (\Delta z \Delta y)\si(x)[1 - (a/2) \exp(- \rho \, n )\sin(k \,n - \phi)] + \mbox{order} \:a^2
\]
from the total second difference.
The final curvature contribution is then
\begin{equation}
    \delta^{(2)}\rE^x_X(LQG;n)/\rE^x_X = (a/2) \,f^2 \,\exp(- \rho \, n ) \sin(k \,n); \: n \neq 0.
\label{DampedCurvature}
\end{equation}
Again, there are no order $a^2$ corrections.

\subsection{Curvature at n = 0}
\label{n=0}
    To dampen the discontinuities at n = 0,
we assume the ratio
$\rho/k$ is small.  This minimizes
the discontinuity in the slope
of the exponent $\exp(-\rho |n|)$ at n = 0, as well as the
discontinuity in the phase $\phi$.  For small $\rho/k$,
\begin{eqnarray}
    \phi &\cong&  + 2 \rho/k \quad (n > 0);\nn
        &\cong& -2 \rho/k \quad (n < 0).
\label{PhiNonZero}
\end{eqnarray}
Since $\rho/\Delta z \sim$ 1/(length of the packet) and
k/$\Delta z \sim$ 1/wavelength, the ratio gives an
estimate of the number of wavelengths
in the central, not strongly damped part of the packet.
\begin{equation}
    \rho/k \sim \mbox{wavelength/(packet length)} << 1.
\label{defr}
\end{equation}
The packet contains many wavelengths.
The relative magnitudes are
\begin{equation}
    \rho \,<< \,k \,<< \,2 \rho/k \,\sim \,\phi.
\label{OrdersOfMagnitude}
\end{equation}

    Because of the discontinuity in $\phi$,
$E^x_X$ at n = 0 is undefined.  We parameterize it
as
\begin{equation}
    \rE^x_X(n=0) = (1 + a_1 + a_2)\Delta x \Delta z,
\label{ExZero}
\end{equation}
where $a_p$ is of order $a^p$ in the small
amplitude a.  The $\rE^x_X$ at n = $\pm 1, \pm2$
follow from \eq{ExponentialPacket}.
\begin{equation}
 \{\rE^x_X(\pm 2) \cong  \rE^x_X(\pm 1)\} = \{1 \pm a (\sin \phi)/2  - (a^2/32) \cos (2 \phi)\}\Delta x \Delta z. \nn
\label{ExAtpmOne}
\end{equation}
We have kept leading
order in the smaller quantities $\rho$ and k,
and (temporarily) all orders in $\phi$.

    The $a_i$ in \eq{ExZero}can be determined by
requiring the order $a^2$ corrections to
curvature to vanish, as at \eq{NoSecondOrder}.
The $a_i$ contribute to curvature only at
n = $\pm $1 and n = 0.
\begin{align}
   \delta^{(2)}\rE^x_X(\pm 1)/\rE^x_X(\pm 1) & =
                \{\rE^x_X(\pm 2) - 2 \rE^x_X(\pm 1) + \rE^x_X( 0)\}/\rE^x_X(\pm 1) \nn
    &=  a_1 +a_2 + (1 + a_1)[\,\mp a(\sin \phi) /2  \,] \nn
   &\quad + a^2[\, \cos (2 \phi)/32 + (1/4)\sin^2 \phi];\nn
   \delta^{(2)}\rE^x_X(0)/\rE^x_X(0) & =
                \{\rE^x_X(+1) - 2 \rE^x_X(0) + \rE^x_X( -1)\}/\rE^x_X(0) \nn
                &=-a^2/16 \cos (2 \phi) - (a_1+ a_2) + 2(a_1)^2,
\label{NearZeroCurvature}
\end{align}
to order $a^2$.
Setting $a^2$ curvature terms to zero gives
\begin{eqnarray}
    a_1 &=& 0;\nn
    a_2 &=& - a^2 /32,
\label{ai}
\end{eqnarray}
neglecting terms of second order in $\phi$.  The surviving contributions
to curvature are now
\begin{eqnarray}
    \delta^{(2)} \,\rE^x_X/\rE^x_X(\pm 1) &=&  \mp a(\sin \phi)/2;\nn
     \delta^{(2)} \,\rE^x_X/\rE^x_X(0) &=& 0.
\label{CurvatureNearZero}
\end{eqnarray}

    The discontinuities in the triads are also minimized.  Compare
\begin{eqnarray}
    \rE^x_X(0)  &=& 1 - a^2/32; \nn
    \rE^x_X(\pm 1)  &\cong& 1 \mp (a/2) \sin(k-\phi) -(a^2/32).
\label{Compare}
\end{eqnarray}
In a theory so fundamentally discrete as LQG, some traditional
notions of continuity may have to be abandoned; however, the
above values for curvature and triad establish a smooth
extrapolation through n = 0.

\subsection{The ADM energy}

    The expression \eq{ExponentialPacket} contains undamped terms
involving
\[
    2 \,\rho \,\mid n \mid -1.
\]
The quantum triads diverge at infinity.

    Initially, these divergent terms were included to make the
solution analytic in $\rho$, in the limit $\rho \,\rta \,0$.  With
these terms included, the
damped form also reduces correctly to the undamped
form, \eq{QuSineWave}.

    These terms also have a fundamental
significance, however.  Because the rest of \eq{ExponentialPacket}
is damped, these are the only terms which survive at
large $\mid \mathrm{n} \mid$, therefore the only terms which
contribute to the surface term in the Hamiltonian.
Some terms must survive, or the ADM energy will vanish.
The ADM energy is computed in appendix \ref{ADM}.

\section{Coherent states}\label{Coherent}

    This problem requires both U(1) coherent states
(for longitudinal holonomies, along z) and SU(2) coherent
states (for transverse holonomies, along x and y).  This
difference (U(1) \emph{vs.} SU(2))
is a consequence of the intial gauge fixing which reduces the
full 3+1 dimensional problem to the planar problem.
The connection reduces to 1 x 1 and 2 x 2 subblocks \cite{HusSmo}.
\[
    \mathrm{A}^{Z}_{x,y} = \mathrm{A}^{X,Y}_z = 0.
\]
Longitudinal holonomies
\[
    \exp[i \int \mathrm{A}^{Z}_z \,\rS_Z]
\]
involve only \A{Z}{z} and
are U(1) rotations around Z.  Transverse holonomies
\[
    \exp [i \int (\mathrm{A}^{X}_a \rS_X + \mathrm{A}^{Y}_a \rS_Y )], \quad a = x,y
\]
involve no $\rS_Z$ (the axis of rotation lies in the XY plane) but
are otherwise full SU(2) rotations.

    Longitudinal coherent states
are parameterized by a peak rotation angle
and its conjugate variable, the component of angular
momentum along z. The longitudinal coherent states
have been constructed elsewhere \cite{GCSIII},
and will not be discussed here.

    Construction of the transverse, SU(2) coherent
states required an entire paper \cite{1}.
However, the basic structure of these states
should not be surprising to anyone familiar
with coherent states for a free particle.
The next subsection reviews construction of the free particle
coherent states.  A follow-on section reviews
the construction of the SU(2) coherent states,
emphasizing the close parallel between the free particle and SU(2) cases.

\subsection{The free particle analogy}
\label{FreeParticle}

    The recipe for constructing a
coherent state for the free particle starts from a wave function
which is a delta function.
\[
    \delta (x-x_0) = \int exp \,[ \,ik(x-x_0)] \, dk/2\pi.
\]
This wave function is certainly strongly peaked, but it is not
normalizable. Also, it is peaked in position, but it needs to be
peaked in both momentum and position.  To make the packet
normalizable, insert a Gaussian operator $\exp (-p^2/2\sigma^2)$.
(Choosing the Gaussian form is a "cheat", because we know the
answer; but for future reference note that all the eigenvalues
$k^2$ of $p^2$ must be
positive, so that the Gaussian damps for all k.) To produce a
peak in momentum, complexify the peak position: $x_0 \rightarrow
x_0 + i p_0 /\sigma ^2$.  With these changes, the packet becomes

\begin{equation}
\begin{split}
    \rN\int  \exp [-p^2/(2\sigma^2)] &\exp[i k(x-x_0) + k p_0/\sigma^2] \, dk/2\pi\\
        &=  \rN\int \exp [-k^2/(2\sigma^2) + i k(x-x_0) + k p_0/\sigma^2]\, dk/2\pi\\
        &=(\rN \exp (p_0^2/2 \sigma^2)/\sqrt{2\pi}) \cdot \exp[ip_0(x-x_0) -(x-x_0)^2 \sigma^2/2].
\end{split}
\label{freecoh}
\end{equation}
The last line, which follows after completing the square on the
exponential, exhibits the characteristic coherent state form.

    There is not just one coherent state, but a family of coherent
states, characterized by the parameter $\sigma$.  The shape of the
wave function is highly sensitive to $\sigma$; but the peak values
($x_0, p_0$) are independent of $\sigma$, as is the minimal
uncertainty relation $\Delta x \Delta p = \hbar /2$.  The
coherent states constructed below contain a parameter t which is
analogous to $1/\sigma^2$.

    Now apply the above recipe to the SU(2) planar case.
The free particle states are parameterized by peak values of
two conjugate variables (x,p), whereas the SU(2) states
are parameterized by peak values of conjugate angles and
angular momentum.  Both the conjugate variables may be thought
of as vectors, since the angles determine the rotation
vector for the holonomy
(directed along the axis of rotation, with magnitude the angle
of rotation).
Because angles are peaked, the  holonomies are peaked.
Because angular momentum is peaked, the \Etld
are peaked.

    The first step in the recipe requires construction of a delta
function (in angle, since angle is the new coordinate replacing position x).  One might start
from the simplest holonomy, which is
\begin{eqnarray}
        h^{(1/2)} &=&\exp[ \,i \,\hat{m} \cdot \vec{\sigma}\,\theta/2 \,] \nn
                 &=&  h^{(1/2)}(-\phi+\pi/2,\theta,\phi-\pi/2); \nn
    \hat{m} &=& (\cos \phi, \sin \phi,0).
        \label{Defm}
 \end{eqnarray}
$h^{(1/2)}$ has rotation axis along $\hat{m}$,
magnitude of rotation $\theta$, and angular momentum 1/2.  A hat denotes a unit vector.
$\hat{m}$ has no component along z because the gauge fixing has
eliminated the \A{Z}{x,y}.  The middle line is the usual Euler angle decomposition.
A complete set of rotation matrices on the group manifold (LQG) replaces
the complete set of plane waves on the real line (free particle).
The matrices have the same Euler angle decomposition as the
simplest holonomy.
\begin{align}
    \delta (\theta - \alpha)\delta(\phi - \beta)/\sin (\alpha)
           &=  \sum_{J,M}((2J+1)/4\pi) \rD^{(J)}(h)_{0M} \rD^{(J)}(u)_{0M}^* ;\nn
    \rD(h) &= \rD(-\phi+\pi/2,\theta,\phi-\pi/2);\nn
    \rD(u) &= \rD(-\beta+\pi/2,\alpha,\beta-\pi/2); \quad \mbox{(wrong)}.
\label{wrongdelta}
\end{align}
Proceeding along these lines, one would arrive at states very
similar to those constructed by Thiemann and Winkler
for the general case of full local SU(2) symmetry \cite{GCSIII,GCSI,GCSII}.

    As indicated on the last line of \eq{wrongdelta},
this is not the correct procedure.
It fails to take into account the holonomy-flux algebra, which
produces an anticommutator.
\begin{equation}
    \{\mathrm{E}^a_A, h_a\} = i (\gamma\kappa/2)[\sigma_A/2, h_a]_{+}.
\label{Anticommutator}
\end{equation}
The \Etld are double grasp: they grasp both incoming and
outgoing holonomies at the vertex.  Because the transverse topology is $\mathrm{S}_1$,
one and the same holonomy is both incoming and outgoing.  It is
grasped on both sides, leading to the anticommutator.  (In the usual 3 + 1 case,
the holonomy connects two different vertices.  Even if an \Etld is
double grasp, the \Etld can grasp only one side of a holonomy.)

    The anticommutator maps the three
matrix elements of h into themselves.  (There are only
three independent elements of h, not four.  Because the axis of
rotation lies in the XY plane, the
two diagonal elements of h are equal.)
The action of the \Etld on the three h is
isomorphic to the action of the generators of the
rotation group O(3) on the three dimensional
representation of O(3), the spherical harmonic $\rY^M_L$ with L = 1.

    In fact the matrix elements of h are proportional to
spherical harmonics, although spherical harmonics with
unusual angular dependence.  From \eq{Defm},
\begin{eqnarray*}
      (\mathcal{N}/\sqrt{2}) h_{\mp,\pm} &=&  \mp \mathcal{N} \,\sin(\theta/2)\exp[\pm (i\phi - i \pi/2)]/\sqrt{2}\nn
                                    &=& \,Y^{\pm}_1(\theta/2,\phi - \pi/2); \nn
         \mathcal{N} h_{++} = \mathcal{N} h_{--} &=& \mathcal{N} \cos (\theta/2) \nn
                           & =& \rY^{0}_1(\theta/2,\phi - \pi/2) .
\end{eqnarray*}
The subscripts on h abbreviate the spin values; e.~g. $h_{+-}$
is the element in row $m = +1/2$ and column $m = -1/2$.

    It is
interesting that the h are proportional to spherical harmonics;
but the  essential feature is that anticommutator $[\sigma_a/2, h]_{+}$
maps h $\rta$ h in the same way that the O(3) generator $\rS_a$
maps $\rY_1 \rta \rY_1$.  For a sample anticommutator calculation
which illustrates this mapping,
see appendix \ref{HilbertSpace}.  The unconventional
half-angle $\theta /2$ is a reminder that
the Y's are constructed from holonomies $h^{(1/2)}$ depending
on a half-angle.

    To obtain the higher spin representations
of the O(3) symmetry, one couples together L = 1 representations
in the usual manner to form the L $>$ 1 representations $\rY^M_{L}$.
The action of the \Etld is given by
the matrix elements of an O(3) generator.
\begin{eqnarray}
    (\gamma\kappa/2)^{-1}\rE^x_{\pm} \,\rY^M_{L} &=&
            \Sigma_{N} \rY_{L N} \,\bra{L,N} \mathrm{S}_{\pm}\ket{ L,M}; \nn
             f_{\pm} &:=& (f_x \pm i f_y)/\sqrt{2}.
\label{Ytransf}
 \end{eqnarray}
$\rY_{L M} = \rY_{L M}(\theta /2,\phi -\pi /2)$.  For L = 1
the $\rY_{1 M}$ reduce to matrix elements of h, and
the anticommutator gives the expansion on the right-hand
side of \eq{Ytransf}.

    \Eq{Ytransf} gives the two \Etld isomorphic to
$\mathrm{S}_{\pm}$.  What is the operator isomorphic
to $\mathrm{S}_0$?  It cannot be \E{x}{Z} since that
field has been gauged to zero.  If one applies the commutator of the
\E{x}{\pm} to h, one finds an operator
\begin{equation}
    (\gamma\kappa/2)^{-1}\rE^a_0 h_a := [ \,h_a, \sigma_z/2 \,]_-  .
\label{defE0}
\end{equation}
Note the commutator.  One can verify directly that this
commutator is isomorphic to the action of $\rS_0$: the
diagonal elements of h (isomorphic to $\rY_0$) are mapped into zero;
off-diagonal elements (isomorphic to $\rY_{\pm}$)
are multiplied by factors of $\pm $1/2.  The action on a general Y
is \eq{Ytransf} with $\mathrm{S}_{\pm}$ replaced by
$\mathrm{S}_0$.
\begin{equation}
     (\gamma\kappa/2)^{-1}\rE^x_0 \,\rY^M_{L} =
              M \,\rY^M_L  .
\label{YtransfE0}
\end{equation}

    When the O(3) symmetry is taken into account,
the correct formula for the delta function is
\begin{eqnarray}
    \delta (\theta/2 - \alpha/2)\delta(\phi - \beta)/\sin (\alpha/2)
        &=& \sum_{L,M} \rY_{L M}(h) \rY_{L M}(u)^*; \nn
        \rY(h) &:=& \rY(\theta/2, \phi - \pi/2); \nn
        \rY(u) &:=& \rY(\alpha/2, \beta - \pi/2).
\label{rightdelta}
\end{eqnarray}
u is the peak value of h.

    u is a representation of both the original SU(2) gauge
group and the new O(3).  Within SU(2), $u^{1/2}$ is the
peak value of $h^{1/2}$ with angle of rotation $\alpha$.  When the SU(2) matrix elements are
regrouped into  L = 1 representations of O(3), u is
the peak value for the L=1 representation, with angle of rotation $\alpha/2$.
In a coherent state context, u is usually the O(3) u.

    The delta function may also be expressed in terms of
rotation matrices, since Y is just a rotation matrix.
\begin{equation}
    \rY_{LM}(u) = \sqrt{(2\rL+1)/4\pi}\: \rD^{(L)}_{0M}(-\beta+\pi/2,\alpha/2,\beta-\pi/2).
\label{YEqD}
\end{equation}
The axis of rotation for u must lie in  the xy plane,
since u is the peak value of h.  This dictates the Euler angle decomposition.

    Continue with the recipe for constructing the coherent
state: dampen the sum using a Gaussian
\[
    \exp [-t \,\rL(\rL+1)/2].
\]
The parameter t is the analog of the parameter $1/\sigma^2$
in the free particle case.
Complexify by extending the angles in u to complex
values, replacing u by a matrix g in the complex extension
of O(3).  The coherent state has the general form
\begin{equation}
    \ket{ u,\vec{p}} = \rN \sum_{L,M}
                    \exp[-t \rL(\rL+1)/2] \rY(h)_{LM} \rY(g)_{LM}^*.
\label{defcoh}
\end{equation}

    Every matrix g in the complex extension of
O(3) can be decomposed into a product
of a Hermitean matrix times a unitary matrix (''polar
decomposition''; see for example \cite{Halltext}).
\begin{equation}
    g =  \mbox{ Hermitean x unitary}.
\label{polar}
\end{equation}
The complexification in the free particle
case is also a product of factors.  The "matrices"
in that case are 1 x 1.
\[
    \exp[-i k x_0] \rta \exp[-i kx_0 + k p_0/\sigma^2].
\]
Here, $\exp[-i k x_0]$ plays the role of the unitary factor.  The
free particle analogy suggests that the
Hermitean factor should contain a vector related to (angular) momentum.

    There are a lot of matrices in the complex
extension.  Some trial-and-error is needed to obtain
the desired peak properties.  The natural first choice for
the unitary factor in \eq{polar} is u,
the value of g in the limit Hermitean matrix $\rta$ 1.  This choice
leads to the simplest proofs.

    The Hermitean factor (:=$\mathcal{H}$) may be
parameterized by a vector $\vec{p} = p\, \hat{p}$.
In the fundamental representation,
\begin{equation}
    g = \mathcal{H} \,u = \exp\, (\,\vec{\sigma}\cdot \vec{p}/2 \,) \,u.
\label{defp}
\end{equation}
The vector $\vec{p}$ gives the matrix $\mathcal{H}$ an axis
$\hat{p}$, analogous to axes $\hat{m}$ and $\hat{n}$
for matrices h and u.

    The higher order representations $\rY_{LM}$
may be complexified similarly.
\begin{eqnarray}
    \rD^{(L)}_{0M}(u)\,&\rta& \exp [\vec{\rS}^{(L)}\cdot \vec{p} \,]_{0R} \,\rD^{(L)}_{RM}(u) \nn
        & = & \mathcal{H}^{(L)}\, u^{(L)}
        \label{geqHu}
\end{eqnarray}
This formula replaces the Y's by the corresponding rotation matrices,
in order to clarify the matrix multiplication.

    $\mathcal{H}^{(L)}$ is expected to diverge as $\exp (p\,\mathrm{L})$
for large L, because of  its  $\exp [\vec{\rS}\cdot \vec{p} \,]$
form.  Combine this with the damping factor:
\begin{equation}
    \exp[-t\,\mathrm{\rL(\rL+1)}/2\,]\,\exp[\, p\,\mathrm{L}\,] = \exp\{-(t/2)[\rL + 1/2 - p/t]^2 + f(t,p)\}.
\label{peakL}
\end{equation}
The exponent has a maximum at an $<\rL>$ given by
\[
     <\mathrm{L}> + 1/2 =  p/t.
\]
The 1/2 looks a bit peculiar until one realizes
\[
    \sqrt{\rL(\rL+1)} \cong \rL + 1/2.
\]
Evidently the coherent states tend to maximize
$\sqrt{\rL(\rL+1)}$ rather than L.  Usually the 1/2
will be dropped.

    All three axes of rotation are assumed to lie in the xy plane:
$\hat{p}, \,\hat{m},$ and $\hat{n}$   for $\mathcal{H}$, h, and u respectively.
\begin{eqnarray}
        \hat{p} &=& (\cos (\beta + \mu), \sin (\beta + \mu), 0);\nn
        \hat{m} &=& (\cos \phi, \sin \phi, 0); \nn
        \hat{n} &=& (\cos \beta, \sin \beta, 0).
\label{defnp}
\end{eqnarray}
$\mu$ is the angle between the peak axis of rotation $\hat{n}$ and $\hat{p}$.
Of course the axis of u should lie in the XY plane, because
u is the peak value of h, and the axis of h is in the XY plane.

    Placing  $\hat{p}$ in xy plane is a bit worrysome, because it seems
to suggest the angular momentum is restricted to the xy plane.
However, we shall see in the next section that the angular momentum
is not $\hat{p} $ but rather $\hat{p}$ rotated through u.

\subsection{Basic matrix elements}\label{MatrixElements}
\begin{eqnarray}
    (2/\gamma\kappa)\rE^a_A(LQG)\ket{ u(n),\vec{p}(n)} &=& <\mathrm{L}(n)> \,\hat{p}_B \, \rD^{(1)}(u)_{BA}\ket{ u,\vec{p}} + \mathrm{SC};\nn
    <\mathrm{L}(n)> &=& \,p(n)/t;\nn
    \mathbf{\hat{h}}\ket{ u,\vec{p}} &=&  \rmi \,\mbox{\boldmath$\sigma$}\cdot \hat{n} \sin(\beta/2)\ket{ u,\vec{p}} + \mathrm{SC};\nn
    \mathbf{\bar{h}}\ket{ u,\vec{p}}    & =&  \mathbf{1}\cos(\beta/2)\ket{ u,\vec{p}} + \mathrm{SC}.
\label{peakEh}
\end{eqnarray}
There are two transverse directions, a = x,y.  Therefore
each of the above equations is actually two equations,
one for x and one for y.  The brackets around $<\rL>$ are
of course designed to distinguish the peak value from
the variable L which is summed over in e.~g.~ \eq{rightdelta}.

    The direction of angular momentum is given
by a rotated version of $\hat{p}$ (first line).  SC denotes the small correction
states.  The last two lines give
the matrix elements for the two parts of the holonomy.
\[
    \mathbf{h} = \mathbf{\bar{h}} + \mathbf{\hat{h}}
\]
Only $\mathbf{\hat{h}}$ occurs in the small sine Hamiltonian.
The explicit dependence of $\hat{p}_B \, \rD^{(1)}(u)_{BA}$
on the angles $\mu, \alpha, \beta$ will be derived
at  a later point, \eq{Generalp} of section \ref{FixParameters}.
For completeness, here is the result.
\begin{eqnarray*}
   \hat{P}^a_A(\alpha) := \hat{p}^a_B \,\rD^{(1)}(u_a)_{BA} &=& \cos \mu_a \,\hat{n}_a \nn
        &&\quad  + \sin \mu_a [\cos(\alpha_a/2)\hat{Z} \times \hat{n}_a\,+\sin(\alpha_a/2)\hat{Z}].\; \nn
     \hat{n}_a &=& (\cos \beta_a, \sin \beta_a, 0).
\end{eqnarray*}
u, the peak value of the holonomy, is a rotation through $\alpha/2$
around the axis $\hat{n}$.
$\mu$ is the angle between  $\hat{n}$ and $\hat{p}$.

    There are now two $\hat{p}$ vectors.  The original $\hat{p}$,
introduced at \eq{defp}, characterizes the complex extension of O(3),
and lies in the XY plane.  The new $\hat{P}$, just introduced,
is the original $\hat{p}$, after a rotation by u. The new $\hat{P}$  gives the
direction of angular momentum, from \eq{peakEh}.

    Longitudinal matrix elements resemble the transverse ones.
\begin{eqnarray}
    (2/\gamma\kappa)\rE^z_Z(LQG)\ket{ <\theta_z>,<m_Z>} &=& <m_Z>\ket{ <\theta_z>,<m_z>};\nn
     \mathbf{\hat{h}_z}\ket{ <\theta_z>,<m_z>}&=&  \rmi\mbox{\boldmath$\sigma$}_z \sin(<\theta_z>/2)\ket{ <\theta_z>,<m_z>};\nn
     \mathbf{\bar{h}_z}\ket{ <\theta_z>,<m_z>} &=& \mathbf{1}\cos(<\theta_z>/2)\ket{ <\theta_z>,<m_z>}.
\label{peakEhz}
\end{eqnarray}
Again, each  \eq{peakEhz} is really a pair of equations.  If the holonomy
is outgoing (respectively, incoming), then the peak angle is labeled
$\theta_z(n,n+1)$ (respectively, $\theta_z(n-1,n)$)),
and peak z component of angular momentum is $m_f$ (respectively, $m_i$).

    Table~\ref{t1} lists the various parameters occurring in the
coherent state, together with a brief definition.  Occasionally, where there is
no danger of confusion, the parameters will be written without
their characteristic transverse label a = x or y.
\begin{table}[ht]
\caption{\label{t1}Parameters occurring in the coherent state.  a = x or y}
\begin{ruledtabular}
\begin{tabular}{ll}
parameter& definition\\ \hline
$u_a $ &peak value of transverse SU(2) holonomy $h^{1/2}$; its matrix \\
        & elements also form an L = 1 representation of O(3).\\
$\hat{n}_a$ & axis of rotation for $u_a$; lies in XY plane.\\
$\alpha_a, \beta_a$ & $u^a = u^a(\alpha_a, \beta_a)$ ; $\alpha_a$ = angle of rotation around $\hat{n}_a$;\\
                &$\beta_a $= angle between $\hat{n}_a$ and X axis.\\
$\rD^{(L)}(u)_{M0}$& O(3) rotation matrix with same $\hat{n}$ as $u_a$, but\\
            & rotation angle $\alpha_a/2$; product of  L copies \\
                &of $u_a$, considered as L = 1 rep of O(3).\\
$\vec{p}^a$&vector in XY plane characterizing the complex rotation\\
                & $\exp[\vec{\rS}\cdot\vec{p}]$ multiplying each term in the \\
                &coherent superpostion; $\vec{p}^a = p^a \hat{p}^a$.\\
$\vec{P}^a(\alpha_a)$ &$ \vec{p}^a_B \,\rD^{(1)}(u_a)_{BA} $; $\vec{p}^a$ after rotation by u;\\
                & gives direction of  angular momentum $\vec{\rL}^a$.\\
$\mu_a $& angle between $\hat{p}^a$ and $\hat{n}^a$. \\
$\rM^a$& peak value of Z component of transverse angular momentum.\\
$m_f, m_i$& peak value of Z component of angular momentum, carried by\\
            &Z axis holonomies entering ($m_i$) or leaving $(m_f)$ vertex n.\\
$<\theta_z>$ & peak value of angle for the Z axis holonomy.\\
\end{tabular}
\end{ruledtabular}
\end{table}

    Despite the use
of an O(3) basis, we do not lose information
about SU(2).
\[
   \rD(-\phi+\pi/2,\theta,\phi-\pi/2)^{(j)} \,\ket{ u,\vec{p}} = \rD(-\beta+\pi/2,\alpha,\beta-\pi/2)^{(j)} \,\ket{ u,\vec{p}},
\]
where $ \rD^{(j)}$ is a representation of SU(2).

\subsection{The $\Delta x^i$ should be simple}

    Evidently the LQG formulas for the triads will contain
factors of $\Delta x^i$.  These parameters are largely arbitrary,
and to keep formulas simple, We choose them to be positive and
independent  of $n_z$.  We respect the symmetry by choosing
\[
    \Delta x = \Delta y.
\]

      Global and local Lorentz coordinates $x^i$ and $X^I$
are related by
\[
    x^i =  X^I e^i_I = X^I \rE^i_I /\mid e \mid .
\]
If the two coordinates increase in opposite directions, then the corresponding
\E{i}{I}(cl) has leading term -1 and $\Delta X^i/ \Delta x^i$ is negative.
Since the $\Delta x^i$ have been chosen always positive,
\begin{eqnarray}
    \Delta X^I &=& |\Delta X^I|\, \si(i); \nn
    \rE^i_I(FT) & =&  \si(i) + \cdots.
\label{defsgni}
\end{eqnarray}
sgn(i) is the sign of \E{i}{I} and $e^i_I$.

    In the present gauge ($e^Z_z = \pm 1$),
$\Delta Z = \pm \Delta z$.
Only the $\Delta X, \Delta Y$ can vary with $n_z$;
$\Delta Z$ is a constant.

    Although the formulae of this paper are worked out for
both signs of the sgn(i), this is overkill.  One may always
choose sgn(i) = +1, and the sign does not
change in the small sine limit.  Working out
the results for both signs does help in checking the
algebra.  Section \ref{Sign} and the conclusion
summarize results for the choice sgn(i) = +1.

\section{Determining the coherent state parameters}
\label{FixParameters}
    The Hamiltonian, \eq{H4}, is correct
for one specific set of gauge conditions and constraints.   However,
the coherent states just constructed above are  general. They are
neither gauge-fixed nor constrained.  Imposition of constraints
and gauges determines the peak values u and $\vec{p}$ .

    The states must obey nine constraints: four
single polarization constraints (which constrain the four off-diagonal
transverse \Etld and transverse K to vanish);
two unidirectional constraints; two diffeomorphism constraints;
and the Gauss constraint.

    A coherent state "obeys" a constraint when the
peak values satisfy the constraint.  The
state is usually not an eigenfunction of the
constraint.

    The Hamiltonian depends only on \Etld; but the
constraints depend on extrinsic curvature K and
spin connection $\Gamma$ as well.  The
next section relates K and $\Gamma$ to the
basic quantities \Etld and $\hat{h}$.

\subsection{$\rK$ and $\Gamma$}

    From paper I, the connection A becomes -2 $\rmi \,\hat{h}^I_i$
in the small sine (SS) limit.  K becomes
\begin{eqnarray}
    \gamma \rK^I &=& \rA^I - \Gamma^I \:\mbox{(FT)}\nn
    & \rta & -2i \,\hat{h}^I - \Gamma^I \:\mbox{(SS)}\nn
    & =& 2 \,\hat{n}^I \sin (\alpha/2) - \Gamma^I.
\label{defK}
\end{eqnarray}
The last line expresses $\hat{h}$ in terms of the
peak values for the angle of rotation $\alpha$, and
axis of rotation
\[
    \hat{n}= (\cos \beta, \,\sin \beta,\,0).
\]
For longitudinal fields, the $\hat{h}(n)$ on the
second line of \eq{defK} is replaced by the
average of the two z holonomies at vertex n:
\[
    \hat{h}_z(n) := [\, \hat{h}_z(n,n+1) + \hat{h}_z(n-1,n) \,]/2,
\]
where $\hat{h}$(n,n+1) is the holonomy on edge (n,n+1).

    Now consider $\Gamma$.  From \cite{Semiclassical}
the products $\Gamma \cdot$ E are given by
\begin{eqnarray}
    \Gamma^Y_x \mathrm{E}^x_X + \Gamma^X_y \mathrm{E}^y_Y &=&  [\cd \mathrm{E}^y_Y/\mathrm{E}^y_Y - \cd \mathrm{E}^x_X/\mathrm{E}^x_X] \mathrm{E}^z_Z; \nn
    \Gamma^Y_x \mathrm{E}^x_X - \Gamma^X_y \mathrm{E}^y_Y &=& \cd \mathrm{E}^z_Z.
\label{GammaEqTriads}
\end{eqnarray}
In the present gauge  we may use \eq{EliminateEzZ} to replace
$\cd \mathrm{E}^z_Z$ on the last line by
\[
     [\cd \mathrm{E}^x_X /\mathrm{E}^x_X + \cd \mathrm{E}^y_Y /\mathrm{E}^y_Y]\mathrm{E}^z_Z,
\]
then solve for the individual $\Gamma \cdot$ E.
\begin{eqnarray}
    \Gamma^X_y \mathrm{E}^y_Y &=& - \cd (\mathrm{E}^x_X) \, \mathrm{E}^y_Y/\rC(LQG) ;\nn
    \Gamma^Y_x \mathrm{E}^x_X &=& + \cd (\mathrm{E}^y_Y) \, \mathrm{E}^x_X/\rC(LQG); \nn
    \rC(LQG) &=& \si(e) \, (\Delta z)^2.
\label{GammaInGauge}
\end{eqnarray}
The two $\Gamma$ in \eq{GammaInGauge} are the only ones which occur in the
constraints.  The single polarization constraints force all other $\Gamma$
to vanish.

    All non-basic variables (K, $\Gamma$) are now expressed  in terms of
basic variables ($\hat{h}$, \Etld).  The latter in turn
have been expressed in terms of coherent state parameters at
\eq{peakEh}.

 \subsection{Evaluation of the \mbox{\boldmath $\beta_a$}}

    It is a bit easier to work with the combinations $\mathrm{U}_1 \pm \mathrm{U}_3$
of unidirectional constraints from paper I.  Using \eq{defK}, the K's may be replaced by
combinations of the \Etld and holonomies, quantities with known action on coherent states.
\begin{eqnarray}
    0 &=& [ \,\mathrm{K}^Y_y \mathrm{E}^y_Y + \mathrm{E}^z_Z \,\cd (\mathrm{E}^x_X)/\mathrm{E}^x_X \,]/\sqrt{\mathrm{E}^z_Z} \nn
    &=& \{ \,2 \sin \beta_y \sin(\alpha_y/2)/\gamma + \cd (\mathrm{E}^x_X)/\rC(LQG) \,\}\mathrm{E}^y_Y/\sqrt{\mathrm{E}^z_Z};\nn
    0 &=& [ \,\mathrm{K}^X_x \mathrm{E}^x_X + \mathrm{E}^z_Z \,\cd (\mathrm{E}^y_Y)/\mathrm{E}^y_Y\,]/\sqrt{\mathrm{E}^z_Z} \nn
     &=&  [ \,2 \cos \beta_x \sin(\alpha_x/2)/\gamma + \cd (\mathrm{E}^y_Y)/\rC(LQG)\,]\mathrm{E}^x_X /\sqrt{\mathrm{E}^z_Z}.
\label{UniCombos}
\end{eqnarray}

    Similarly, the  single polarization constraints $\mathrm{K}^X_y = \mathrm{K}^Y_x$ = 0
may be expressed in terms of the \Etld and holonomies, using \eqs{defK}{GammaInGauge}.
\begin{eqnarray}
     0& =& \gamma \mathrm{K}^X_y \nn
    &=& 2 \cos \beta_y \sin(\alpha_y/2) + \cd (\mathrm{E}^x_X)/\rC(LQG); \nn
    0 &=& \gamma \mathrm{K}^Y_x \nn
    &=& 2 \sin \beta_x \sin(\alpha_x/2) - \cd (\mathrm{E}^y_Y)/\rC(LQG).
\label{SPConstraints}
\end{eqnarray}
The unidirectional constraints have an additional \Etld/$\sqrt{\mathrm{E}^z_Z}$
on the right.  However, this additional factor merely produces
a constant, when acting on a coherent state.  Therefore this
factor may be commuted to the left.  The two sets of
constraints, unidirectional and single polarization,
agree only if
\begin{eqnarray}
    (\cos \beta_x) &=& - \gamma\sin \beta_x; \nn
    \cos \beta_y &=& + (\sin \beta_y)/\gamma; \nn
    \cos \beta_x &=& \si(\hat{n}_x) \gamma/\sqrt{1 + \gamma^2}; \nn
    \sin \beta_x &=& - \si(\hat{n}_x) 1/\sqrt{1 + \gamma^2};\nn
    \cos \beta_y &=& \si(\hat{n}_y)/\sqrt{1 + \gamma^2};\nn
    \sin \beta_y &=& \si(\hat{n}_y)\gamma/\sqrt{1 + \gamma^2},
\label{Beta}
\end{eqnarray}
where $\si(\hat{n}_a) = \pm 1$, and $\hat{n}_a$ is the axis of
rotation for the peak holonomy $u_a$.  There is a
sign ambiguity because the first two lines determine $\beta_a$
only mod $\pi$, where $\beta_a$ is
the angle $\hat{n}_a$ makes with the X axis.  Therefore
$\hat{n}_a$ is determined only up to an
overall sign (equivalently, only up to a reflection through
the origin).  For any choice of signs,
the two rotation axes $\hat{n}_x$
and $\hat{n}_y$ are 90 degrees apart.

    The unidirectional and single polarization constraints
are now equivalent.  One can drop the unidirectional
constraints and focus on the single polarization constraints;
the number of independent constraints has dropped to seven.

\subsection{Evaluation of the \mbox{\boldmath $\mu_a$}}

    The single polarization constraints also
require vanishing of off-diagonal angular momentum components.
\[
    \mathrm{E}^x_Y = \mathrm{E}^y_X = 0.
\]
Translated into coherent state language, this
means the off-diagonal components of the
rotated vector $\hat{P}$ must vanish:
\[
    \hat{P}^x_Y(\alpha_x) = \hat{P}^y_X(\alpha_y) = 0.
\]

    The single polarization constraints force \emph{both}
rotated $\hat{P}^a$ and unrotated $ \hat{p}^a$ to depend on
on angle of rotation $\alpha_a$.  For example, if
$\alpha_x =0$, then $\rD^{(1)}(u_x)$ is the identity matrix, in
\eq{peakEh}) and $\hat{P}^x$ reduces to $\hat{p}^x$.
Single polarization therefore requires $\hat{p}(\alpha_x=0)$
to lie along $\pm \hat{X}$.  This determines
$\mu_x(\alpha_x=0)$ in \eq{defnp}.  Note the $\beta$ are
fixed, from the discussion in the preceding section.

    As $\alpha_x$ increases from zero,
$\hat{p}$ must move away from the X axis ($\mu$ must change),
so that the larger rotation
can rotate a larger Y component of $\hat{p}$ onto the Z axis.

    In more detail, the components of $\hat{p}$, both rotated
and unrotated, and for arbitrary polarization, are
\begin{eqnarray}
    \hat{p}^a &=& \cos \mu_a \,\hat{n}_a + \sin \mu_a \,\hat{Z}\times \hat{n}^a ; \nn
    \hat{P}^a &=& \cos \mu_a \,\hat{n}
        + \sin \mu_a [\cos(\alpha_a/2)\hat{Z} \times \hat{n}\,+\sin(\alpha_a/2)\hat{Z}].
\label{Generalp}
\end{eqnarray}
Proof: The unrotated $\hat{p}$ lies in the xy plane, therefore
has components along $\hat{n}^a$ (rotation axis for u,
so also in xy plane) and $\hat{Z}\times \hat{n}^a $.
The angle between $\hat{p}$ and $\hat{n}$
is $\mu$, which gives the first line of \eq{Generalp}.  After
$\hat{p}$ is
rotated through $\alpha/2$ around axis $\hat{n}$, the
angle between $\hat{n}$ and $\hat{P}$ remains $\mu$, which
explains the $\hat{n}$ term on the second line.
After rotation, the vector $\hat{Z}\times \hat{n}^a$,
perpendicular to the axis of rotation, becomes the square
bracket on the last line. $\Box$

    We set components $\hat{P}^x_Y$ = $ \hat{P}^y_X$ = 0, keeping in mind
\begin{eqnarray*}
    \hat{n} &=& (\cos \beta, \sin \beta,0);\nn
    \hat{Z}\times \hat{n} &=& (-\sin \beta, +\cos \beta, 0).
\end{eqnarray*}
Then
\begin{eqnarray}
    \hat{P}^x_Y(\alpha_x) &=& \cos \mu_x \sin \beta_x + \sin \mu_x \cos(\alpha_x/2)\cos \beta_x = 0; \nn
    \hat{P}^y_X(\alpha_y) &=& \cos \mu_y \cos \beta_y - \sin \mu_y \cos(\alpha_y/2)\sin \beta_y = 0.
\label{OffDiagP}
\end{eqnarray}
Because of these relations, the surviving on-diagonal
components simplify greatly.
\begin{eqnarray}
    \hat{P}^x_X &=&  \cos \mu_x /\cos \beta_x; \nn
    \hat{P}^y_Y &=&  \cos \mu_y /\sin \beta_y.
\label{E=0Constraints}
\end{eqnarray}

    \Eq{OffDiagP} may be solved for the $\mu$ in terms
of the (still unknown) $\cos (\alpha/2)$,  using \eq{Beta}.
\begin{equation}
    \tan \mu_a = + 1/[\gamma \cos(\alpha_a/2)].
\label{Mu}
\end{equation}
$\alpha_a$ is assumed small (semiclassical limit:
$\alpha_a$ near 0 = flat space).  Division by $\cos (\alpha/2)$
is therefore legal.  For $\alpha_a = 0$  (no rotation) \eq{Mu} predicts
$\hat{P}^a = \pm \hat{A}$, as expected from the earlier, qualitative
discussion.

    For $\alpha_a \neq 0, \mu$ is constant to order $a^2$.
To show this we expand
\[
    \mu(\alpha) = \mu(0) + \Delta \mu,
\]
and also expand $\cos(\alpha/2)$ in powers of $\sin(\alpha/2)$.
The result is
\begin{equation}
    \Delta \mu =  [\gamma/(1+ \gamma^2)]\sin^2(\alpha/2)/2.
\label{Deltamu}
\end{equation}
From the next section, the sine is order a, therefore
$\mu$ is constant to order $a^2$.

    \Eq{Mu} determines the angle ($\mu_a$)  mod $\pi$.
\begin{eqnarray}
    \sin \mu_a &=& \si(\hat{p}_a) /\sqrt{1 + \gamma^2 \cos^2(\alpha_a/2)}; \nn
    \cos \mu_a &=& \si(\hat{p}_a) \gamma \cos(\alpha_a/2)/\sqrt{1 + \gamma^2 \cos^2(\alpha_a/2)},
\label{SignMu}
\end{eqnarray}
$\si(\hat{p}_a) = \pm 1$.
$\hat{p}_a$, like $\hat{n}$, is determined only up to a reflection through
the origin.


    \Eq{E=0Constraints} determines the
magnitude of the transverse components of $\hat{P}^a$ and relates the
direction of $\hat{P}^a$  to signs defined earlier.
\begin{equation}
    \hat{P}^a_A(\alpha_a) = \si(\vec{p}_a)\si(\vec{n}_a)[1 + \Or a^2]; \quad a = A.
\label{sgnp}
\end{equation}
$\hat{P}^a_A$ is approximately a unit vector, $\pm 1$. A way to understand the
sign: sgn(vector) = +1 means a small angle (first or fourth quadrant),
sgn(vector) = -1 means add $\pi$; the angle $\hat{P}$ makes with the x-axis
is therefore near $\pi$ ($\hat{P} \cong -1)$ when sgn$(\hat{p})$ = - sgn$(\hat{n})$.

    \Eq{sgnp} allows a Z component of order a.  $\hat{P}^a_Z$ follows from \eq{Generalp},
\[
    \hat{P}^a_Z(\alpha_a) =  \sin \mu_a \sin(\alpha_a/2).
\]
From the next section,  $ \sin(\alpha_a/2) = \Or$a.

    The unidirectional constraints
and two of the single  polarization constraints are now satisfied.
The remaining constraints are the two single polarization constraints,
two diffeomorphism constraints, and Gauss.

\subsection{Determination of \mbox{\boldmath $\sin(\alpha/2)$}}

    Section \ref{Damping} constructed a set of transverse \Etld which
satisfy the scalar constraint.  One can insert those \Etld into the remaining single polarization
constraints \eq{SPConstraints}, and thereby determine $\sin(\alpha/2)$,
$\alpha/2$ the peak angle of rotation.
\begin{eqnarray}
    0 &=& \gamma\mathrm{K}^B_a, \: a \neq B,\nn
     &=& 2 \sin \beta_x \sin(\alpha_x/2) - \cd \mathrm{E}^y_Y(cl)\,(\Delta x/\Delta z)\,\si(e) \nn
     &=& 2 \cos \beta_y \sin(\alpha_y/2) + \cd \mathrm{E}^x_X(cl)\,(\Delta y/\Delta z )\,\si(e)
\label{EqnForAlpha}
\end{eqnarray}
From \eq{ExponentialPacket},
\begin{eqnarray}
    \cd \mathrm{E}^x_X(cl;n)&=&  \si (x)\{- (a/2) f \exp(- \rho\, n )\cos(k \,n - \phi /2)  \nn
               &&\qquad  +(-a^2/16) [f \exp(- 2\rho \, n ) \,\cos(2 k \,n - 3 \phi/2) \nn
                &&\qquad + [- \rho \,\exp(- 2\rho\, n ) + \rho ] \,(f/\rho^2) \, \cos \phi]\}; \nn
                 f^2 = (k^2+\rho^2).
\label{DeltaE}
\end{eqnarray}
Insert \eqs{DeltaE}{Beta} into \eq{EqnForAlpha}.
\begin{align}
 - 2\, \si(\hat{n}_x) \sin(\alpha_x/2) &= + \sqrt{1 + \gamma^2}\,\cd \mathrm{E}^y_Y(cl)(\Delta x/\Delta z) \,\si(e) \nn
                          &= \si(x) (a/2) f \exp(- \rho\, n )\cos(k \,n - \phi /2)(\Delta x/\Delta z) \,\si(e) + \Or a^2;\nn
   2\, \si(\hat{n}_y) \sin(\alpha_y/2)&= - \sqrt{1 + \gamma^2}\,\cd \mathrm{E}^x_X(cl)\,(\Delta y/\Delta z )\,\si(e) \nn
                             &= + \si(y)(a/2) f \exp(- \rho\, n )\cos(k \,n - \phi /2)\,(\Delta y/\Delta z) \,\si(e) + \Or a^2.
\label{Alpha}
\end{align}
As advertised, $\sin(\alpha_x/2)$ is order a.

\subsection{$<\rL(n)>$}

    The first \eq{peakEh} contains a magnitude factor $<\rL(n)>$,
which determines fluctuations along the direction of $\vec{\rL}$,
times a unit vector which determines fluctuations perpendicular
to the direction of $\vec{\rL}$.
\[
     (2/\gamma\kappa)\rE^a_A(LQG)  =  <\mathrm{L}(n)> \,\hat{p}_B \, \rD^{(1)}(u)_{BA}
\]
Insert \eq{ExponentialPacket} for \Etld on the left, and
use \eq{sgnp} to simplify the $\hat{P}$ factor.
\[
     (2/\gamma\kappa)[1 - E_1 + E_2]\Delta y \Delta z \,\si(x) = <\mathrm{L_x}(n)> \,\si(\hat{p}_x) \si(\hat{n}_x)[1 + \Or a^2].
\]
To avoid distracting detail, the classical field theory part of
$\rE^a_A(LQG) = \rE^a_A(FT) \Delta y \Delta z  $
is written as a series of terms $E_p$ of order $a^p$.
Since $<\rL_a>$ is positive, we must set
\begin{equation}
    \si(a) = \si(\hat{p}_a) \si(\hat{n}_a), \quad a = x,y.
\label{sgn=sgn}
\end{equation}
$<\rL_a>$ has the same n dependence as \E{a}{A}(FT).
\begin{eqnarray}
    <\rL_x>(n) & =& \rL_{0x} [1 - E_1 + E_2];\nn
    \rL_{0x} &=& (2/\gamma\kappa)\Delta y \Delta z.
\label{Ln}
\end{eqnarray}
The equation for $<\rL_x>$ has opposite sign for the $E_1$ term,
but the same value for $\rL_0$, because of the  choice
$\Delta x = \Delta y$.  One can drop the x subscript
on $\rL_{0x}$.

    One may choose $<\rL_a>$ any convenient
size by adjusting the $\Delta x_a$.  \Eq{Ln}
is a precise statement of this scaling behavior:  the
amplitude  $\rL_0$ scales with the $\Delta x_a$.

\subsection{\mbox{\boldmath $\mathrm{K}^Z_z$}, \mbox{\boldmath $\mathrm{E}^z_Z$} and Gauss}

    The four single polarization constraints are now satisfied. Gauss and
two diffeomorphism constraints remain.

    The diffeomorphism constraints from paper 1 are (for p = 1/2)
\begin{eqnarray*}
    1 &=& \Etwo/(\mathrm{C(LQG)}\mathrm{E}^z_Z); \nn
    \mathrm{C(LQG)} &=&  (\Delta z)^2 \,\si(e) ; \nn
      0 &=&  \mathrm{K}_z  .
\end{eqnarray*}
The last line yields
\begin{align}
    0 &= \gamma\mathrm{K}^Z_z(n) \nn
    &= -2i \,[\,\hat{h}^Z_z(n,n+1) + \hat{h}^Z_z(n-1,n)\,]/2 - \Gamma^Z_z \nn
    &= 2 \,[\,\sin(<\theta_z>/2)(n,n+1) + \sin(<\theta_z>/2)(n-1,n)\,]/2 -0 .
\label{KzEqZero}
\end{align}
Either all peak $\theta_z$ are zero, or $\theta_z$ alternates between
two values having opposite sign.  Since holonomic angles should go to
zero in the flat space in front of the packet, $<\theta_z>$
= 0.

    The remaining diffeomorphism constraint may be used to show  \E{z}{Z},
\Etwo, and $(m_f + m_i)$ are constants, to order $a^2$.
Since \E{z}{Z} grasps on both sides of the vertex,
its expectation value depends on $m_f + m_i$, the expectation
values of $\mathrm{S}_z$ on the ingoing plus outgoing sides of
the vertex.
\begin{equation}
     \mathrm{E}^z_Z(LQG) = (\kappa\gamma/2) (m_f + m_i) .
\label{mfPlusmi}
\end{equation}
The LQG values are related to classical values by the diffeomorphism
constraint \eq{QuGaugeChoice}.
\[
      \mathrm{E}^z_Z(LQG) =(\rE^x_X \, \rE^y_Y)(LQG)/\rC(LQG)
             =  \,(\rE^x_X \, \rE^y_Y)(cl)\, \si(e) \,\Delta x  \Delta y .
\]
The classical \Etld have the form
\begin{eqnarray*}
    \rE^x_X (FT) &=& (1 - A_1 + A_2) \si(x);\\
    \rE^y_Y(FT) &=& (1 + A_1 + A_2) \si(y),
\end{eqnarray*}
where $A_p = \Or a^p$.  Therefore
\begin{equation}
     \mathrm{E}^z_Z(LQG) = \si(z) \,[\,1 - A_1^2 + 2 A_2 + \Or a^3\,] \Delta x  \Delta y.
\label{EzIsConstant}
\end{equation}
Comparison of \eqs{mfPlusmi}{EzIsConstant} gives
\begin{equation}
    (2/\kappa\gamma) \mathrm{E}^z_Z(LQG) = (m_f + m_i) = \si(z) (2/\kappa\gamma)\,\Delta x  \Delta y (1 + \Or a^2).
\label{mfPlusmiIsConstant}
\end{equation}
\E{z}{Z}(FT), \Etwo (FT), and $(m_f + m_i)$ are constants, to order $a^2$. $\Box$

    \Eq{mfPlusmiIsConstant} is another example of scaling
behavior.  The overall amplitudes (but not the fluctuating
factors) scale with the $\Delta x^i$.

    The quantity $m_f - m_i$ occurs in Gauss' Law.
Gauss requires a vanishing net flow of z momentum through all
six sides of the cube surrounding a given vertex.  Equivalently,
if the $\beta_x, \,\beta_y, \,\mbox{and}\, \theta_z$ at a given vertex
are all subjected to the same rotation, the product of
holonomies at the vertex must be invariant.  This requires
\[
    M_x + M_y + (m_f - m_i) = 0.
\]
The first two terms are the net outflow of Z angular momentum contributed by the transverse
directions; the last parenthesis is net outflow contributed by the
z holonomies.

    The expectation value of  $M_a$ is given by the
operator \E{a}{0}, \eqs{peakEh}{Generalp}.
\begin{eqnarray}
    M_a &=& \mathrm{<L_a>} \,\hat{p}^a_B \, \rD^{(1)}(u_a)_{B0} \nn
    &=& \mathrm{<L_a>} \,\sin \mu_a \sin(\alpha_a/2).
\label{peakM}
\end{eqnarray}
Gauss then requires
\begin{equation}
    0 = <\rL_x> \,\sin \mu_x \sin(\alpha_x/2) + <\rL_y> \,\sin \mu_y \sin(\alpha_y/2) + (m_f-m_i).
\label{GaussMSum}
\end{equation}
From \eq{Alpha} $\sin (\alpha_x/2)$ is a power series in a of the
form
\begin{equation}
    \sin (\alpha_x/2) = \si(x) \si(\hat{n}_x) [-A_1 + A_2 + \cdots]\Delta x/\Delta z,
\label{PowersinAlpha}
\end{equation}
where $A_p = \Or a^p$.
From \eq{Ln},
\begin{equation}
    <\rL_x> = L_0 [1 - E_1 + E_2 + \cdots].
\label{PowersinAlpha2}
\end{equation}
 From \eq{Mu}, $\mu$ depends on $\cos (\alpha/2)$,
therefore the power series for $\sin \mu$ goes as
\begin{equation}
    \sin(\mu_x) = \si(\hat{p}_x)[B_0 + B_2 + B_3 + \cdots],
\label{PowersinMu}
\end{equation}
where $B_p = \Or a^p$.

    Now insert these expansions into \eq{GaussMSum} (as usual,
changing  the sign of odd powers of a for the y term).
The product of sign factors in each
term equals +1, because of \eq{sgn=sgn}.
\Eq{GaussMSum} then collapses to (for $\Delta x = \Delta y$)
\begin{equation}
    2(\Delta x /\Delta z) [B_0 A_2 + E_1 A_1 + \Or a^3 \,]\rL_0 + (m_f - m_i) = 0.
\label{mfMinusmiIsConstant}
\end{equation}
From $\rL_0 \sim \Delta x \Delta z)$ plus \eq{mfPlusmiIsConstant},
\begin{equation}
    m_f = m_i = \si(z) (\kappa\gamma)^{-1}\,(\Delta x)^2 (1 + \Or a^2).
\label{m}
\end{equation}

    As a check on calculations we use formulas for $\hat{P}$ to
evaluate the diffeomorphism constraint
\[
    (\Delta z)^2 \rE^z_Z (LQG) = \si(e) \,\Etwo (LQG).
\]
Acting on a coherent state, this becomes
\begin{align}
 (\Delta z)^2 \rE^z_Z(FT) \Delta x  \Delta y  &= \si(e) (\kappa\gamma/2)^2 (<\rL_x \rL_y>) \hat{P}^x \times \hat{P}^y \cdot \hat{Z}; \quad \mbox{or} \nn
 (\Delta z)^2 \si(z)[1 + \Or a^2] &=\si(e)( \Delta z)^2 \si(x) \si(y)[\,1 + \Or a^2 \,].
\label{GaugeCheck}
\end{align}
From the analysis at \eq{mfMinusmiIsConstant}, the \E{z}{Z}
are order unity plus order $a^2$, which explains the square
bracket on the left, second line. From \eqs{sgnp}{sgn=sgn}
\[
     \hat{P}^a_A  = \si(a)[1 + \Or a^2],
\]
which explains the square bracket on the right, second line.
This check
shows that the formulas for $\hat{P}$ are consistent with
the diffeomorphism gauge.

    In the above calculation \eq{GaussMSum} was used for Gauss, rather than its
small sine approximation,
\[
    0 = \cd \rE^z_Z + (-2i) \hat{h}^A_a \rE^a_A.
\]
The latter is not quite as accurate.  For example,
$\cd  \rE^z_Z = m_f - m_i$,  but only after using
slow variation.  The small sine version is fine
when Gauss occurs multiplied by factors of sine, as
in the Hamiltonian and vector constraints.  When
Gauss is stand-alone, \eq{GaussMSum}
is more accurate.

\section{Coordinate oscillations}

    $\Delta Z $ does not oscillate, since the $\Delta x^i$
are fixed, and
\begin{equation}
    \Delta Z = e^Z_z \Delta z = \si(z) \Delta z.
\end{equation}
The oscillations in transverse coordinates follow
from \eqs{areaEqArea2}{ExponentialPacket}.
\begin{eqnarray}
    \Delta Y &=& \si(y) \Delta y \{\, 1 -(a/2) \,\exp( \mp \rho \, n ) \,\sin[k\,  n\mp \phi] + \Or a^2\,\}; \nn
    \Delta X &=& \si(x) \Delta x \{\, 1 +(a/2) \,\exp( \mp \rho \, n ) \,\sin[k\,  n\mp \phi] + \Or a^2\,\}.
\label{TransverseOscillations}
\end{eqnarray}

\section{The metric at spatial infinity}

    To this point the calculation has been carried out
to order $a^2$ in the small amplitude
a.  This is fine, except for the undamped part of the amplitude,
which diverges at infinity.
\begin{eqnarray}
    \rE^a_A(LQG) &=& (\Delta z \Delta x^b)\si(x) \{1+ \cdots \nn
                &&+ \,(a^2/32)(\mp 2 \rho \,n  +1) \,(f/\rho^2)  \,\cos \phi\},
\label{undamped}
\end{eqnarray}
from \eq{ExponentialPacket}.  z = $\pm \mid z \mid$.
If there are divergent corrections of higher order in n,
they will be needed to compute the ADM energy.

    It is safe to assume that the
space outside the wavepacket is flat.  The
present solution is time varying, and the wave has not
yet reached spatial infinity, which must be flat therefore.
In flat space both
the scalar constraint and the Riemann tensor must vanish.
From \eq{WeylEinsteinEDoubleDot},
\begin{equation}
   \partial^2_u \rE^x_X = \partial^2_u \rE^y_Y = 0.
\label{FlatMetric}
\end{equation}
The variable $\sqrt{2}$ u = (z-ct) corresponds to the
discrete variable n.  In the present small sine LQG
approach, derivatives with respect to $\sqrt{2}$ u
become differences with respect to n.  The \Etld are
therefore linear functions of n at infinity.  \Eq{ExponentialPacket}
for  the \Etld diverges linearly at infinity,
therefore is correct as it stands.  There are no higher
order corrections in n (though there may be higher
order corrections in a).

    The surviving terms at $n \rta \pm \infty$
may be read off from \eq{ExponentialPacket}.
\begin{eqnarray}
    \rE^x_X = \rE^y_Y &\rta& (\Delta z \Delta y)\si(x) \{1+ \,(-a^2/32) \,(  \pm 2 \rho \,n  -1) \,(f^2/\rho^2)  \,\cos \phi\} \nn
    &:=& (\Delta z \Delta y)\si(x) \{1 \pm D \, n + D_0\}.
    \label{defDD0}
\end{eqnarray}
$D_0$ and D are constants of order $a^2$.  If terms down by
$\rho/k \ll 1$ are dropped,
\begin{eqnarray}
        D_0 &=&  (a^2/32)(f/\rho^2)  \,\cos \phi \cong 1 + (a^2/32)(k/\rho)^2;\nn
        D \, n &  \cong&  -(a^2/16)(k^2/\rho) n.
\label{DandD0}
\end{eqnarray}

    \E{z}{Z} follows from the gauge choice \E{z}{Z} $\propto$ \E{x}{X}\E{y}{Y},
\eq{QuGaugeChoice}.
\begin{eqnarray}
    \rE^z_Z(LQG)& =& \si(e) (\Delta z \Delta y)[1 \pm D \, n + D_0]^2 \nn
                & \cong & \si(e) (\Delta z \Delta y)[1 \pm 2 D \, n + 2 D_0];\nn
    \cd \rE^z_Z(LQG)& \cong & \si(e) (\Delta z \Delta y)[ \pm 2 D ].
\label{EzatInfinity}
\end{eqnarray}
The second and third lines drop order $a^4$ terms, which are inaccurate
because the \E{a}{A} are known only to order $a^2$.

    Since $\cd$ \E{z}{Z} appears in the surface term for the energy,
it is useful to check the above result by a second method: solve
the constraint \Ht = 0. From paper I, section on the final form
of the Hamiltonian,
\begin{eqnarray}
    \mathrm{\Nt\Ht +ST}& \rta & \sum_n (1/\kappa)\{\cdots \nn
        & & + \cd \mathrm{E}^z_Z \,[- (\cd \Etwo)/\Etwo \nn
        & & + \cd \mathrm{E}^z_Z/2\mathrm{E}^z_Z ]+ \cd(\cd \E{Z}{z}) \} = 0.
\label{H3}
\end{eqnarray}
We have dropped a term proportional to
\[
    (\cd \mathrm{E}^y_Y /\mathrm{E}^y_Y  -\cd \mathrm{E}^x_X /\mathrm{E}^x_X)^2 \rta \, 0.
\]
This expression vanishes at infinity for the present explicit solution,
and also generally, because it represents the non-gauge physical
degree of freedom, which should be absent in flat space.
We use the diffeomorphism gauge \eq{QuGaugeChoice} to replace
\begin{eqnarray*}
    (\cd \Etwo)/\Etwo &=& \cd \mathrm{E}^z_Z/\mathrm{E}^z_Z  \\
                    &=& \cd \mathrm{E}^y_Y /\mathrm{E}^y_Y  + \cd \mathrm{E}^x_X /\mathrm{E}^x_X \\
                    &\rta& 2 \cd \mathrm{E}^x_X /\mathrm{E}^x_X .
\end{eqnarray*}
\Eq{H3} becomes
\[
    -\cd \mathrm{E}^x_X /\mathrm{E}^x_X + \cd(\cd \E{Z}{z})/\cd \mathrm{E}^z_Z = 0.
\]
The solution is
\begin{equation}
    \cd \E{Z}{z} \rta A \mathrm{E}^x_X,
\label{cdEz}
\end{equation}
A a constant.  We know \E{x}{X} is linear in n at infinity,
from the argument at \eq{FlatMetric}.  We now
know that $\cd $\E{Z}{z} is also linear in n.
\Eq{cdEz} agrees with our previous result for
$\cd$\E{Z}{z} if we take A = $\pm $ 2D, use \eq{defDD0} for
\E{x}{X}, and drop order $a^4$.

\section{Signs}
\label{Sign}

 \subsection{More information about sgn($\hat{n}_a$)}

        We now know the vectors $\hat{P}_a $ and $ \hat{n}_a$
(unit vectors along the angular momenta and axes of rotation, respectively)
from \eqs{Beta}{SignMu}, but only mod $\pm 1$.  However,
we have some limited information about the sign of $\hat{n}_a$.
From \eq{Alpha}, the leading order a contributions to $\sin(\alpha_x/2)$
and $\sin(\alpha_y/2)$ have the same magnitude, but may differ in sign.
The n-dependent  factors cancel out of the ratio of sines.
\begin{eqnarray}
    \cos \beta_x \sin(\alpha_x/2)/\sin \beta_y \sin(\alpha_y/2)& =&  - \si(x)/\si(y); \quad \mbox{or} \nn
    \si(\hat{n}_x) \sin(\alpha_x/2)/\si(\hat{n}_y) \sin(\alpha_y/2) & =&  - \si(x)/\si(y).
\label{SignbetaSalpha}
\end{eqnarray}
\Eq{SignbetaSalpha} determines the relative sign of the $\hat{n}_a$, but only if
we know the relative sign of the $\sin(\alpha_a/2)$!

   It turns out only the
signs of the \emph{products} ($\cos\beta_a$ or $\sin\beta_a$ or \si($\hat{n}_a$))
times $\sin(\alpha_a/2)$
are significant.  The basic holonomy is
\begin{eqnarray}
    h_a &=& \cos (\alpha_a/2) + i \, \sigma\cdot \hat{n}_a \sin(\alpha_a/2) \nn
    &=& \cos (\alpha_a/2) + i \,\sigma\cdot (\cos \beta_a,\sin \beta_a, 0)\sin(\alpha_a/2).
\label{sinsgnSignificant}
\end{eqnarray}
This expression is invariant under simultaneous sign change of both $\hat{n}$
and $\alpha$.  I.~e. a rotation through $\alpha $ around axis
$\hat{n}$ is equivalent to a rotation through $-\alpha$ around $-\hat{n}$.
If a solution exists for one sign
of $\hat{n}$ and $\alpha$, then an identical solution exists
for the opposite sign of $\hat{n}$, provided we simultaneously change the
sign of $\alpha$.  The signs of  $\alpha_a$ and $\hat{n}_a$ have little
physical significance when considered separately, and their products are
constrained only by \eq{SignbetaSalpha}.

\subsection{Signs, for sgn(i) = +1}

    It is useful to examine the pattern of signs
for the simplest and most natural case: axes $x^i$ and $X^I$
running in the same direction; right-handed coordinate system:
sgn(i) = sgn(e) = +1.  For this case,
the two angular momenta $\rL_a \,\hat{P}^a$ have their
largest component along the positive $\hat{A}$ direction,
from \eqs{sgnp}{sgn=sgn}.
\begin{equation}
    \hat{P}^a_A = (+1)\hat{A} + \Or a^2; \quad a = A.
\label{hatpforsgn+1}
\end{equation}
\Eq{sgn=sgn} then requires
\begin{equation}
    \si(\hat{P}_a) = \si(\hat{n}_a).
\label{sgnmu=sgnbeta}
\end{equation}

    As explained at \eq{sinsgnSignificant},
only the sign of the product
$\sin(\alpha_a/2)\, \si(\hat{n}_a)$ is physically significant.
These products are constrained by \eq{SignbetaSalpha}, which
allows only two physically distinct sign patterns.
\begin{eqnarray}
    [\sin(\alpha_x/2), \si(\hat{n}_x)] &\Rightarrow& [\sin(\alpha_y/2),\si(\hat{n}_y)]:\nn
      \mbox{pattern 1: } [\pm,\mp] &\Rightarrow&  [\pm,\pm];\nn
      \mbox{pattern 2: } [\pm,\pm] &\Rightarrow&  [\pm,\mp].
\label{SgnPatterns}
\end{eqnarray}
In words: the sign pattern for $[\sin(\alpha_x/2), \si(\hat{n}_x)]$
implies the pattern for $[\sin(\alpha_y/2),\si(\hat{n}_y)]$;
if $\sin(\alpha_x/2)$ and $\si(\hat{n}_x)$ have opposite
sign, then $\sin(\alpha_y/2)$ and $\si(\hat{n}_y)$ must have the
same sign; and conversely.

    Note the two sign choices in each
square bracket of \eq{SgnPatterns} are physically equivalent;
changing the sign of $\alpha$ and simultaneously reversing the direction of
the rotation axis gives the same physical rotation.  Consequently,
there are only two
physically distinct patterns, rather than eight.  A
choice of sign pattern in \eq{SgnPatterns} fixes  the remaining
sign, sgn$(\hat{p})$, because
sgn$(\hat{p})$ = sgn($\hat{n}$) from \eq{sgnmu=sgnbeta}.

   Given the high degree of symmetry between the x and y
directions, one would expect solutions to occur in pairs
differing by x $\rla$ y.  The two patterns,
\eq{SgnPatterns}, form such a pair.  If one changes a $\rta$ - a,
in addition to relabeling x $\rla$ y, then the curvature is unchanged
and the two patterns become physically identical.

    Table \ref{t2} gives the order a behavior of the
dynamical variables, for the choice sign(i) = 1 and
the $[\pm,\mp] \Rightarrow [\pm,\pm]$ pattern in \eq{SgnPatterns}.
This pattern allows four physically equivalent solutions
corresponding to the four
possible sign choices for the pair $(\hat{n}_x, \hat{n}_y)$.
A relabeling x $\rla$ y generates  the
four $[\pm,\pm] \Rightarrow [\pm,\mp]$ solutions.

\begin{table}[ht]
\caption{\label{t2}Variables to order a, for sign(i) = +1. $f^2 = k^2 + \rho^2 \cong k^2;  \rL_0 = (2/\gamma\kappa)\Delta x\Delta z.$}
\begin{ruledtabular}
\begin{tabular}{ccc}
variable& behavior&reference\\
$\rE(FT)^x_X$ -1& -(a/2)\,$\sin(k n - \phi)$ & \eq{ExponentialPacket} \\
$\rE(FT)^y_Y$ - 1 &+(a/2)\,$\sin(kn - \phi)$ & \eq{ExponentialPacket}\\
$<\rL^a_A> - \rL_0$, a=A       & $\rL_0 \, (\rE(FT)^a_A$ -1) &\eq{Ln}\\
$\rE(FT)^z_Z$ - 1&fixed&\eq{EzIsConstant}\\
$\hat{P}^a_A$, a=A  & +1& \eq{hatpforsgn+1}\\
$\si(\hat{n}_x)\sin(\alpha_x/2)$&-(fa/4)\,$\cos(k n - \phi/2)$&\eq{Alpha}\\
$\si(\hat{n}_y)\sin(\alpha_y/2)$&+(fa/4)\,$\cos(k n - \phi/2)$&\eq{Alpha}\\
$\beta_a $& fixed& \eq{Beta} \\
$\mu_a$& fixed & \eq{Muorder2} \\
\si($\hat{n}_a$) \si($\hat{p}_a$) & +1&\eq{sgn=sgn}\\
$\rM^x = \rL_0 \, \hat{p}^x_Z(\alpha_x)$& -($\rL_0$ fa/4)\,$ \cos(k n -\phi/2)$&\eqs{peakM}{SignMu}\\
$\rM^y = \rL_0 \,\hat{p}^y_Z(\alpha_y)$& +($\rL_0$ fa/4)\,$\cos(k n - \phi/2)$&\eqs{peakM}{SignMu}\\
$m_f, m_i$&fixed $>$ 0 &\eq{m}\\
$<\theta_z>$ &  0 & \eq{KzEqZero}
\end{tabular}
\end{ruledtabular}
\end{table}

\section{Discussion}

    The standard formula for the area,
\[
    \mbox{area} = \kappa\gamma \sqrt{j(j+1)},
\]
might suggest that spins are input and areas an output.
However, when deriving the classical limit, it is
perhaps better to think of area as input, and (average or
peak) spin as output.
For example, from \eq{mfPlusmiIsConstant},
\[
    (2/\gamma\kappa)\Delta x \Delta y [1 + \Or a^2 ] = \si(z) (m_f + m_i).
\]
One adjusts the left hand side, until the right hand side is
large enough to be semiclassical. Classical variables
\Etld(FT) can be near unity, even though LQG eigenvalues are far from
unity, because of the area elements in
\[
    \Etld(LQG) = \Etld(FT) \, \Delta x^i \Delta x^jd
\]
Also, because
of the area elements in the LQG triads,
fixing the diffeomorphism gauge fixes $\Delta \mathrm{Z}$,
the linear spacing between vertices.

    In the flat space surrounding the wave, the holonomic angles
$\alpha_a$ go to zero, but not the canonically conjugate angular momenta.
In flat space, both $\rL_0$ and $m_z$ can be large; see \eq{EeqAngularMomenta}.

    The z components of angular momentum $m_i, m_f$ are not
related to the helicity operator for the wave.
The helicity is given by \cite{helicity}
\begin{equation}
    -2 \rmi \sum_n (\rE^+_+ \rK^-_- - \rE^-_- \rK^+_+ ),
\label{helicity}
\end{equation}
where $f_{\pm} = (f_x \pm \rmi f_y)/\sqrt{2}$.
(The above conserved quantity is a volume sum, rather
than the usual surface term.  For a full discussion see reference \cite{helicity};
but note that the transverse sector resembles special relativity
more than general relativity: the variables (K, \Etld) in the transverse sector
are gauge fixed.)  \Eq{helicity} counts +2 $\hbar$
times the number of spin 2 $\rE^+_+$ minus
2 $\hbar$ times the number of spin -2 $\rE^-_-$.
If the E and K are expanded in terms of more familiar fields,
\begin{equation}
    \rE^+_+ = [\rE^x_X + \rmi \rE^x_Y + \rmi (\rE^y_X + \rmi \rE^y_Y)   ]/2,
\end{equation}
etc., one can show that the helicity operator vanishes, as it should.
From the discussion in appendix \ref{ADM},
$m_f + m_i$ is closely related to energy, rather than helicity.



       In weak field geometrodynamics $g_{xx} $ and $g_{yy}$
oscillate 180 degrees out of phase, giving rise to the usual
picture of a gravity wave as an ellipse with
fluctuating major and minor axes.
The two angles $\alpha_x$ and $\alpha_y$ are also
180 degrees out of phase, but only if the two axes $\hat{n}_a$ are chosen
to have the same sign.  They have the same sign if both are chosen
to be as close to the positive X axis as possible (one axis in first, and
one in fourth quadrant); or both are chosen to be as distant
from the axis as possible (one in second, and
one in third quadrant). They can never be in the same quadrant, because
the $\hat{n}_x$ and $\hat{n}_y$
rotation axes are 90 degrees apart, independent of signs.

    However, in LQG the signs of the $\alpha_a$ are less significant,
because one obtains the same rotation by changing the sign of an $\alpha_a$
while simultaneously reversing the direction of the
axis of rotation.  Therefore
both $\alpha_a$ can be in phase, provided  the two $\hat{n}_a$ have opposite
sign.

    Of course LQG also incorporates the usual picture of the gravity wave as an ellipse.
See for example the formulae for coordinate oscillations,
\eq{TransverseOscillations}.

    The behavior of the transverse holonomies is relatively simple.
Each holonomy is characterized by an axis of rotation $\hat{n}_a$ and
an angle of rotation $\alpha_a$.  $\hat{n}_a$ can be reflected through the
origin, but otherwise cannot change: the angle with the X axis,
$\beta_a$, is fixed by the Immirzi  parameter.  Only $\alpha_a$ can oscillate.
The $\beta_a$ must be fixed in order for the unidirectional and single
polarization constraints to agree.  It is not clear what would happen if
one or both of these constraints were removed.

    In contrast to the transverse holonomies,
longitudinal holonomies are trivial: $<\theta_z>$ = 0.
The longitudinal momenta and angles ($m_z$ and $<\theta_z>$) do not oscillate, to
order a.

        Turning from holonomies to fluxes,
the formalism predicts that the peak values of both transverse $\rE^a_A$
have Z components.  This is a bit surprising.  If one imagines a rectangular box of volume
$\Delta$X$\Delta$Y$\Delta$Z surrounding each vertex,  $\rE^a_A$
supposedly gives the area of the side having normal $\hat{A} \neq \hat{Z}$.
Also, the triad $\rE^a_Z$ has been gauged to zero.

    It is a little easier to understand the need for Z components
if one realizes that change in area with normal $\hat{A}$ produces
change in area with normal $\hat{Z}$.
The triad $\rE^x_X$ (for example) changes because the associated area
$\Delta$X$\Delta$Z changes. That area changes because length $\Delta$X
changes.  ($\Delta$Z is gauge-fixed.)  The changes in $\Delta$X in turn
induce changes in the area $\Delta$X$\Delta$Y with normal $\hat{Z}$.

    The requirement of a Z component also seems to be embedded quite
deeply in the basic equations.  One half of the Gauss law,
\[
    \Gamma^Y_x \mathrm{E}^x_X - \Gamma^X_y \mathrm{E}^y_Y = \cd \mathrm{E}^z_Z.
\]
reduces, in the present diffeomorphism gauge, to
\begin{equation}
    \cd\rE^x_X/\rE^x_X + \cd\rE^y_Y/\rE^y_Y = \cd\rE^z_Z/\rE^z_Z.
\label{DiffForm}
\end{equation}
This expression relates transverse triads to the z triad, which is consistent with the
idea that the transverse triads (produce a change in Z area and) have a small Z component.

    Note the first
differences in \eq{DiffForm}.  The Gauss Law can be rewritten as an
integral over the box $\Delta$X$\Delta$Y$\Delta$Z, with oppositely directed
normals on opposite faces.  This implies the net flow of Z area
is given by a difference.

    \Eq{DiffForm} is roughly
\[
    -ka \cos/(1- a \sin) + ka \cos/(1+a \sin) = \Or a^2,
\]
Each term on the left-hand side is order a (each transverse Z component is
order a), but the sum is odd under (a $\leftrightarrow $ -a);
therefore the right-hand side,  $\rE^z_Z$, is order $a^2$, consistent with the result that
$\rE^z_Z$ does not oscillate to order $a^2$.

    The explicit expressions for transverse momentum are consistent with the
foregoing qualitative discussion.  For example $\vec{\rL^y}$ to order a is
\begin{eqnarray}
    \vec{\rL^y} &=& \rL^y(n)\,\hat{P} \nn
                &=& \rL_0 [1 + (a/2) \sin(k n - \phi)] [\hat{Y} + \hat{Z} (fa/4)\cos(k n - \phi/2)] \nn
                &\cong& \hat{Y}\,[ \rL_0 +  \,\rL_0 (a/2) \sin(kn)] + \hat{Z}\, \rL_0 (fa/4)\cos(kn).
\label{LEllipse}
\end{eqnarray}
For simplicity, the last line drops the phases, which
are order $\rho$/k (small; many wavelengths in a packet).
The Y component measures area $\Delta$X$\Delta$Z; hence
the Y component tracks the variation
of $\Delta$X, \eq{TransverseOscillations} which varies as sin.
The Z component tracks the first difference of the $\Delta$X
in area $\Delta$X$\Delta$Y, therefore varies as the first difference
of  sin, namely $f \cos \cong k \cos$.

    \Eq{LEllipse} is consistent with the entries in table \ref{t2}.
The $\rL^y$ entry, second line, comes from the
$<\rL^a_A> - \rL_0$ entry in table \ref{t2}; the
$\hat{Z}$ term, third line, is the $\rM^y$ entry.

    \Eq{LEllipse} predicts that
both magnitude and direction of $\vec{<\rL^y>}$
oscillate in the presence of a gravitational wave.
The tip of the angular momentum vector traces out a small
ellipse with major axis a and minor axis fa/2 $\cong$ ka/2.
The $\rM^x$ and $\rM^y$
oscillations are 180 degrees out of phase
(provided \si(x) = \si(y) as at table \ref{t2}).

    The above results are largely unaffected by spatial diffeomorphisms,
since the  holonomies and \Etld(LQG) are constructed to be invariant.  Even
the classical \E{a}{A}(FT) are largely invariant.  Change occurs only in
order $a^2$.  From \eq{ClGaugeChoice}, the
classical gauge is characterized by a power p.
The following gauge transformation changes p to p$'$.
\begin{equation}
    z' = \int^z [\si(z) \rE^z_Z]^{(1-2p')/4} \rmd z.
\label{GaugeTransf}
\end{equation}
The above integrand, expanded in powers of a, is unity
plus order $a^2$.  Therefore the order a terms in \Etld(FT)
are invariant.   A corollary: the order a oscillations of
$\Delta X,Y$ are invariant.

    Although the LQG holonomies studied here are classical,
the results carry over to the quantum theory because of the
use of coherent states.  From \eq{Classical=Quantum} the
expectation value of the quantum constraint vanishes, if
the classical constraint vanishes.  Also, the expectation
value of a quantum holonomy varies in the same manner as the
corresponding classical holonomy.

    Although this paper used O(3) harmonics $\rY_L$ rather than
SU(2) harmonics, the two have identical
angular behavior, with
\[
    <\rL> \: = \: <2j>.
\]
As a check: for L = 1, the O(3) harmonics are
combinations of j = 1/2 SU(2) harmonics.

\appendix

\section{The ADM Energy}
\label{ADM}

    From \cite{Semiclassical}, the surface
term is given by $-\Nt\,\cd E^z_Z (LQG)/\kappa$.  From \eqs{EzatInfinity}{DandD0}
for \E{z}{Z} at infinity,
\begin{eqnarray}
 \mbox{ADM Energy} &=& -(\Nt/\kappa)\si(z) \Delta x \Delta y\,  (\mp a^2/8)(k^2/\rho) |^{+\infty}_{-\infty} \nn
                &=& (1/\Delta Z \kappa) \Delta x \Delta y \,\si(z) 2 (a^2/8)(k^2/\rho) ,
\label{ADMEnergy}
\end{eqnarray}
where \mbox{\Nt(LQG) = 1/$\Delta Z$}.  $\Delta Z \,\si(z) = \Delta z$ is positive
(\eq{defsgni}).

    The factor of $\Delta Z \,\si(z) = \Delta z$ looks gauge-variant.
However, we can introduce a k(cl) and $\rho$(cl), defined by
\begin{eqnarray}
    \exp[-\rho n] \sin[k n] &=& \exp[ -\rho(cl) n \Delta z] \sin[k(cl) n \Delta z]; \nn
    n \Delta z &=& z;\nn
    \rho(cl) &=& \rho/\Delta z;\nn
    k(cl) &=& k/\Delta z.
\label{defrhocl}
\end{eqnarray}
If we shift to the classical quantities in \eq{ADMEnergy}, the factor
of $\Delta Z \,\si(z) = \Delta z$ disappears.
\begin{equation}
    \mbox{ADM Energy} = \hbar c (\Delta x \Delta y/\kappa) a^2 [k(cl)^2/4  \rho(cl)] .
\label{ADMEnergy2}
\end{equation}
The first parenthesis is dimensionless because we have been giving
$\kappa$ the dimension of length squared.

    The energy should be proportional to the volume occupied by the wave.
After the shift to classical k and $\rho$, the ADM energy contains a factor
\[
    \Delta x \Delta y/\rho(cl) \sim  \Delta x \Delta y\,\mbox{length}.
\]
The above expression is a measure of volume.  Since the packet
is proportional to   $\exp[- \rho(cl) z]$,
$1/\rho(cl)$ is a measure of the length of the packet.

    Rough estimates of the energy give results similar to \eq{ADMEnergy}.  The energy
in weak field approximation is of order
\begin{eqnarray}
    \int (\partial_z \Etld)^2 &\sim& \mbox{(xy area)}\,(\,k(cl)\, a)^2\int dz [ \,\sin (k(cl)\, z) \exp(-\rho(cl) \,z) \,]^2 \nn
    &=& \mbox{(xy area)}\,(k(cl) \,a )^2 [ \,k(cl)^2/(2\rho(cl)) \,] \,\{ \,1/[\,k(cl)^2 - \rho(cl)^2 \,]\} \nn
    &\approx& \mbox{(xy area)}\, (k \,a/ \Delta z)^2 (\Delta z/2 \rho) ,
\label{BackOfEnvelope}
\end{eqnarray}
The last line of \eq{BackOfEnvelope} neglects terms down by $ \rho(cl)/k(cl) $
(= $\rho/k) \ll 1$).
This back-of-the-envelope estimate
contains the same factors as \eq{ADMEnergy}.

    One might suppose the energy is not quantized, because
periodic boundary conditions
were not used, and k  in \eq{ADMEnergy} can
be anything.  However, see the next section.

\section{Spreading of a coherent wave packet}
\label{StrongSpreading}

    The extent of wave packet spreading was estimated
elsewhere \cite{2}; the present appendix modifies that discussion
for the planar case.  We first argue that all packets approach
(non-spreading) simple harmonic oscillator (SHO) packets in the
limit of large quantum numbers.  We then estimate the lifetime
of the LQG packet.

    For minimal spreading, the
spacing between energy levels of the system should be
as constant as possible, resembling the spacing between levels of the
usual oscillator \cite{Glauber, Klauder}.
Suppose, for example, the energy goes as
$\rL^p$, p some power other than linear, L a quantum number.
(For example, the spherical harmonics making up the coherent state
of the earth have energy
going as $\rL^2$.)  The spacing between levels is
\begin{equation}
    \delta \rE = \mbox{const.} p \rL^{p-1} \delta \rL,
\label{NoSpread1}
\end{equation}
which is no longer in SHO form: a constant times the change of an integer.

    Although the factor multiplying the integer is
now a function, rather than a constant,
the variation of this factor
across the packet is very small.
\begin{eqnarray}
    \delta (\mbox{factor})/\mbox{factor} &=& (p-1) \delta \rL/\rL \nn
        &\sim&  (p-1) \sigma(\rL)/\rL.
\label{NoSpread2}
\end{eqnarray}
$\sigma(L) $, the standard deviation of the L values in
the packet, is expected to be $\ll$ L in the classical limit.
All packets approach a SHO packet in the limit of large
quantum numbers.

    The lifetime of the gravitational wave packet is infinite.
The ADM energy equation \eq{ADMEnergy} contains an area which
is quantized, by equation \eq{m}.  Since the quantum number
is an integer (m, rather than $\sqrt{j(j+1)}$), the energy levels have
the SHO form.

    The area $\Delta$X$\Delta$Y, and hence m, will fluctuate
at finite values of n,  when higher
orders in a are included.  The ADM energy is determined
by long-range "tails", however, which extend beyond the packet.
These presumably do not fluctuate.  The higher orders in a
should correct the constants multiplying $\Delta$X$\Delta$Y
at infinity; the constant spacing between levels will change, but
the spacing will remain uniform.

    We have assumed the remaining constants in the energy
(k, $\rho$, a) are not quantized.  These constants
are unlikely to contain hidden
angular momentum dependence, because they occur in expressions
such as
\[
    \cd \rE^x_X/\rE^x_X, \cd \rE^y_Y/\rE^y_Y
\]
which are independent of area.   Presumably these constants
are determined by the matter source; investigation of the
source is beyond the scope of the present work.

    In principle the mechanism of dispersion in LQG differs
from the mechanism in weak field geometrodynamics.
In the latter theory the graviton is a
superposition of plane waves, and the dispersion of the
packet depends on the velocity spread of the waves
in the packet.  Since every wave has the same
velocity c, spread is zero and packet lifetime
is infinite.

    In LQG the solution at each  vertex is a superposition
of z and transverse holonomies.  It is a bit of luck
energies are SHO in the planar case. Otherwise the  holonomic
packets conceivably could spread,
giving the solution a finite lifetime.




\section{The planar Hilbert space} \label{HilbertSpace}

    The x and y holonomies involve only generators
$\mathrm{S}_X, \mathrm{S}_Y$, since the \A{Z}{a}
have been gauged to zero.
Each transverse holonomy $h^{(1/2)}$ therefore
has an axis of rotation with no Z component.
\begin{eqnarray}
        h^{(1/2)}  &=&\exp[ \,i \,\hat{m} \cdot \vec{\sigma}\,\theta/2 \,]; \nn
        \hat{m} &=& (\cos \phi, \sin \phi,0),
        \label{defm}
 \end{eqnarray}
for some angle $\phi$.  There is one holonomy for
each transverse direction x,y; and one $\phi$ for each transverse
direction, $\phi_x$ and $\phi_y$.  Since the two directions are
treated equally, it is sufficient to discuss only the x holonomies;
the subscript x will be suppressed. When expanded out, the spin 1/2
holonomy $h^{(1/2)}$, \eq{defm}, becomes
\begin{equation}
h^{(1/2)}
    =\left[\begin{array}{cc}
        \cos (\theta/2) & i \exp(-i\phi)\sin (\theta/2) \\
        i\exp(+i\phi)\sin (\theta/2)& \cos (\theta/2)
        \end{array}
        \right]
    \label{defh}
\end{equation}
The usual Euler angle decomposition for this rotation is
\begin{eqnarray}
    h^{(1/2)}& =&  \exp[-i\sigma_Z(\phi-\pi/2)/2 \,]\,\exp(i\sigma_Y
    \theta/2) \,\exp[+i\sigma_Z(\phi-\pi/2)/2] \nonumber \\
   & =&  h^{(1/2)}(-\phi+\pi/2,\theta,\phi-\pi/2).
           \label{EulerDecomp}
\end{eqnarray}

    From the discussion at \eq{Anticommutator},  \Etld
produces an anticommutator.
\begin{eqnarray}
    \rE^x_A h^{(1/2)} &=& \rE^x_A \,\exp[ \,i\int \rA^B_x \, \rS_B \, dx \,]\nn
    &=& (\gamma \kappa/2) \, [\,\sigma_A/2,\, h^{(1/2)}]_+,
\label{actionE}
\end{eqnarray}
Fortunately, the anticommutator reshuffles the elements of h in a relatively
simple way.  Introduce the operators \E{x}{\pm}, where as usual
\begin{equation}
    f_{\pm} := (f_x \pm i f_y)/\sqrt{2}.
    \label{defpm}
\end{equation}
The operators \E{x}{\pm} reshuffle the components of h in the
same way that the familiar angular momentum operators $\mathrm{L}_{\pm} $
reshuffle the L = 1 Legendre polynomials $Y^M_1$.  For example,
write out the action of
the anticommutator in \eq{actionE}, for index A = +.
\begin{equation}
 [\sigma_+/2, h^{(1/2)}]_+
        = \sqrt{1/2}\left[\begin{array}{cc}
        i\exp(-i\phi)(\sin \theta/2) & 2 \cos )\theta/2) \\
        0               & i \exp(+i\phi)\sin (\theta/2)
        \end{array}
        \right]
    \label{E+h}
    \end{equation}
Compare this matrix to the original matrix, \eq{defh}.
\E{x}{+} has reshuffled the matrix elements as
\begin{eqnarray}
    (i/\sqrt{2}) \exp(-i\phi)\sin \theta/2 &\rightarrow&  \cos \theta/2 ;\nn
       \cos \theta/2 &\rightarrow& (i\sqrt{2}) \exp(+i\phi)\sin \theta/2 ;\nn
      (i/\sqrt{2}) \exp(+i\phi)\sin \theta/2  &\rightarrow& 0.
\label{E+Shuffle}
\end{eqnarray}
This is isomorphic to the action of the operator $\mathrm{L}_+ $ on the
L = 1 Legendre polynomials.  The isomorphism is
\begin{align}
  \mathrm{L}_{\pm} &\leftrightarrow 2 \,\mathrm{E}^x_{\pm}/\gamma \kappa; \nn
   \mathrm{L}_0 &\leftrightarrow 2 \,\mathrm{E}^x_0/\gamma\kappa; \nn
    \rY^{\pm}_1 &\leftrightarrow (\mathcal{N}/\sqrt{2}) h_{\mp,\pm} \nn
                           &= \mp \mathcal{N} \,\sin(\theta/2)\exp[\pm (i\phi - i \pi/2)]/\sqrt{2}\nn
                           &= Y^{\pm}_1(\theta/2,\phi - \pi/2);\nn
    \rY^{0}_1 &\leftrightarrow \mathcal{N} h_{++} = \mathcal{N} h_{--} \nn
                       &= \mathcal{N} \cos (\theta/2) \nn
                       & = \rY^{0}_1(\theta/2,\phi - \pi/2).
\label{isomorphism}
\end{align}
Because of the half angles, normalization of the Y's requires integrating
$\theta$ from 0 to 2$\pi$.  $\mathcal{N} = \sqrt{4\pi/3}$.

     Because the $Y^M_1(\theta/2,\phi - \pi/2)$
transform more simply than matrix elements of $h^{(1/2)}$ under the action of \Etld, one
obtains a more convenient basis by using O(3) 3J coefficients and products
of $Y_1$'s, rather than SU(2) coefficients and products of $h^{(1/2)}$'s.  The
resultant basis is just the set of spherical harmonics $\rY^M_L (\theta/2,\phi - \pi/2)$
for O(3).

    One can take into account the y edges as well as the
x edges, by constructing two bases, $\rY^{Mx}_{Lx}$ and
$\rY^{My}_{Ly}$ for holonomies along the x and y directions
respectively. These harmonics transform simply under the action
of the \Etld:
 \begin{equation}
    (\gamma\kappa/2)^{-1}\rE^x_P \rY^M_{L} =
            \Sigma_{N} \rY_{L N}\bra{\rL,\rN} \rS_P\ket{ \rL,\rM},
                \label{Ytransf2}
 \end{equation}
where $\rY_{L M} = \rY_{L M}(\theta /2,\phi -\pi /2)$.  The
unconventional half-angle reminds us of the origin of these
objects in a holonomy $h^{(1/2)}$ depending on half-angles.

    The transverse coherent states constructed here do not have
unique values for $\rM_x$
and $\rM_y$.  These states are superpositions of  $\rD^{(La)}_{0Ma}$
matrices (a = x,y); and the superpositions will contain a range of values $M_a$.
(Similarly, coherent states in the longitudinal direction will not have
definite $m_z$.)  The superpositions are sharply peaked at central
values of the M's, however,
so that M-values which violate U(1) are suppressed.

     The relation between h and the $\rY^M_1$
is
\begin{equation}
    \mathcal{N}\mathbf{h}^{(1/2)} = \mathbf{1}\rY^0_1 + i \rY^+_1 \mathbf{S}_- +i \rY^-_1 \mathbf{S}_+ ,
        \label{heqY}
\end{equation}
where boldface denotes a 2x2 matrix. \Eq{heqY}
demonstrates that the Y's are as "complete"
a set as the elements of $h^{1/2}$, since the three independent
elements of $h^{(1/2)}$ can be expressed in
terms of the three $\rY^M_1(\theta/2,\phi - \pi/2)$ .

\section{Renormalization}
\label{Renormalization}

    This section is intended for readers
familiar with a  coarse-graining recipe developed by a number of authors \cite{Friedel} - \cite{Livine}.
Readers who are not familiar but
would like to learn more probably should start with reference \cite{Livine}.

    The present treatment is hardly coarse-grained.  The number of
vertices per cycle, $N_\lambda$, is assumed to be quite large:
$N_\lambda$ times order 100 Planck lengths is the macroscopic wavelength.  In this
appendix we "coarse-grain": N vertices are replaced by a single vertex.
N may be taken very large, but should be much less than $N_\lambda$, so that
after the coarse-graining there are still a large number of vertices
per cycle.

      The coarse-graining method of the references starts by choosing
a "maximal tree".  This is a tree which contains no loops and passes through each of the
N vertices once and only once.  For a general,
three-dimensional lattice, the maximal tree is not unique, and
one is forced to discuss dependence on choice of tree.  In the
present case, the maximal tree is unique; it is just the z axis.

    After the tree is chosen, one can SU(2) gauge transform each
holonomy along the tree to the unit holonomy.  In effect
this collapses the N vertices to a single vertex.  In the present case
each vertex is the endpoint for two loops, one in the x direction
and one in the y direction.  The single vertex therefore has
2N loops, beginning and ending at that vertex.  (In the literature
this is described picturesquely as a flower diagram having 2N petals. )

    The holonomies along the maximal tree are z holonomies peaked at
$\theta_z = 0$.   The holonomies are already unit holonomies, and
no gauge transformations are needed. (When the holonomies are non-trivial,
further transformations are needed after the N vertices collapse to
one vertex.  Those additional transformations also are not needed.)

    The wavefunction at the surviving vertex is quite complex.
It is a product of N "x" SU(2) coherent states formerly at vertices
1, 2, $\cdots$, N; and N "y" SU(2) coherent states.
To estimate the new peak angular momentum, we repeat the calculation
given at \eq{peakL}.  The x coherent state (for example) now
contains an exponential which is a sum over the N loops.
\begin{eqnarray}
    \exp[\cdots] &=&\exp[-t\,\sum_i\rL_i(\rL_i+1)/2 +\sum_i p_i\,\rL_i\,] \nn
    &=& \exp\{-( t/2)\,\sum_{i=1}^N [\rL_i+1/2 - (p_i/t)\,]^2 + f(p_i,t)\}.
\label{CoarseGrainL}
\end{eqnarray}
From \eqs{peakL}{LEllipse}
\begin{equation}
    p_i/t = <\rL_i +1/2> = \rL_0 (1 + \Or a) = (2/\gamma\kappa)\Delta x\Delta z (1 + \Or a).
\label{pOvert}
\end{equation}
I.~e.~, $p_i/t$ is the peak value of $\rL_i + 1/2$ before coarse-graining;
and the exponent in \eq{CoarseGrainL} can be minimized by retaining those peak
values after coarse-graining.

    The new value for peak L is a bit clearer if the exponent on the first line of \eq{CoarseGrainL}
is rewritten (neglecting terms independent of $\rL_i$)
\begin{eqnarray}
    [\cdots] &=& -(\rN t/2)\{(L+1/2)_{rms}^2 -2\sum_i( p_i/ t \rN)[(\rL+1/2)_{rms} + \rL_i + 1/2 - (\rL+1/2)_{rms}]\} ; \nn
    (L+1/2)_{rms}& =& [\sum_i(L_i+1/2)^2/N]^{1/2}.
\label{CoarseGrainL2}
\end{eqnarray}
If we use the same, or nearly the same, value $\Delta x$ for the length of every transverse loop, then
$p_i$ is independent of i, $(\rL +1/2)_{rms}$ is approximately p/t, and
$\rL_i + 1/2 - (\rL+1/2)_{rms}$ may be neglected.
\begin{equation}
      [\cdots] = -(\rN t/2)\{(L+1/2)_{rms} -\sum_i( p_i/ t \rN)\}^2 + g(p_i,t).
\label{CoarseGrainL2}
\end{equation}
The rms value is peaked; and the peak value is an average over the $p_i$.

    One can also compute the new curvature, which goes as $\ddot{\rE}/\rE$,
double dot denoting a second difference.  Now numerator and denominator
of $\ddot{\rE}/\rE$ become a sum of terms, one from each vertex.  For simplicity
we suppress the index x or y, and write $\rE^{(k)}_i$ for the order
$a^k$ contribution from vertex i.
\begin{eqnarray}
    \ddot{\rE}/\rE &=& \sum_{i=1}^N[\ddot{\rE}^{(1)}_i + \ddot{\rE}^{(2)}_i]/\sum_{i=1}^N (1 + \rE^{(1)})\nn
    &=&\sum_{i=1}^N[\ddot{\rE}^{(1)}_i(1 + \rE^{(1)})/\sum_{i=1}^N (1 + \rE^{(1)})\nn
    &=&\bar{\ddot{\rE}} + \sum_{j=1}^N (\ddot{\rE}_j- \bar{\ddot{\rE}})(\rE_j - \bar{\rE})/\rN(1+\bar{\rE}) \nn
    &=& \bar{\ddot{\rE}} + (-k^2)\sum_{j=1}^N (\rE_j - \bar{\rE})^2/\rN(1+\bar{\rE}),
\label{CoarseCurvature1}
\end{eqnarray}
where the bar denotes an average over N,
\begin{equation}
    \bar{f} := \sum_{i=1}^N f_i/N.
\label{defbar}
\end{equation}
Superscripts (1), (2) have been dropped; all fields are now
$\rE^{(1)}$ fields,
order unity in a.

    The averages may be estimated by replacing sums by
integrals, for example
\begin{eqnarray}
    \bar{\rE} &= &(1/\rN) \sum_{j=1}^N \rE_j \Delta n \nn
    &\cong & (1/\rN) \int_{n_0}^n (- a/2) \sin(kn)\, dn \nn
    & = & (1/\rN) (a/2k) [\cos(kn) - \cos(kn_0)] \nn
    &=& (1/\rN) (- a/k)\sin[k(n+n_0)/2] \sin[k(n-n_0)/2],
\label{CoarseCurvature2}
\end{eqnarray}
where n = $n_0 + \rN$.  For simplicity we have ignored the damping factor.

    Note
we are averaging only over part of one cycle (N $\ll \rN_{\lambda}$)
so that the averages over sinusoids are order unity, not
negligible (as they would be if we were averaging over several
cycles).  In particular,
\[
    \bar{\rE} = \Or (a/(k)(1/ \rN) = \Or (\rN_\lambda/\rN)(a/2 \pi).
\]
Nevertheless we may drop the final sum in \eq{CoarseCurvature1}.
It is order $k^2 a^2$ whereas the $\bar{\ddot{\rE}}$ term is
order $k^2 a$.

    For the curvature we expect
\begin{equation}
    \ddot{\rE}/\rE = -(k^2 a\zeta/2) \sin[k(n+n_0)/2].
\label{CurvatureAnsatz}
\end{equation}
I.~e., the sine is evaluated in the middle of the interval ($n,n_0$)
and the factor $\zeta$ takes into account the possibility of a
renormalization of the amplitude a.  Comparing
\eqs{CoarseCurvature2}{CurvatureAnsatz}, we have
\[
    (1/\rN) (- ka) \sin[k(n-n_0)/2] = -(k^2 a\zeta/2).
\]
With $(n-n_0)$ = N, k = $2\pi/\rN_\lambda$, this gives
\begin{equation}
    \zeta = \sin(\pi \rN/\rN_\lambda)/(\pi \rN/\rN_\lambda).
\label{Renormalization}
\end{equation}
For N $\rta$ 1 (the smallest possible value of N) N/$\rN_\lambda \cong 0$,
and $\zeta$ has the correct limit $\zeta \rta 1$.

\section{The U(N) formalism}
\label{UN}

    A number of authors have developed a formalism which avoids
explicit SU(2) rotation matrices and uses a
representation of SU(2) based on holomorphic functions
(Bargmann representation)\cite{LQGasSHO,FreidelCoherent,Tambornino, U(N)}.
The approach involves a number of
operators which together form a representation of U(N);
we will refer to this approach as the U(N) approach.  In this
appendix we assume the reader is already somewhat familiar with
the U(N) formalism; readers who desire an introduction might try
reference \cite{U(N)}.

    The U(N) approach shifts the focus from holonomy on edge
e to spinors located at the two ends of edge e.  In particular the
holonomy on the transverse x edge is replaced by two spinors,
a source spinor at the beginning of the edge, and a target
spinor at the end:

\begin{gather*}
    \begin{pmatrix}s_+ \\ s_- \end{pmatrix}, \quad
    \begin{pmatrix}t_+ \\ t_- \end{pmatrix}.
\end{gather*}
The U(N) formalism works with spinor operators
as well as spinor peak values, when  a coherent state basis is used.
To be clear, the above spinors are the peak values.  We suppress
edge labels (x, y, or z) on the spinors.  Until further notice we
consider only x spinors.

    In order to express the peak spinors in terms of
the parameters used in the present paper,
we associate each spinor with a vector according to the following theorem.
The spinor $\chi$,
\begin{gather*}
   \chi(\xi,p_z) =  \begin{pmatrix} \sqrt{1+p_z} \\ \sqrt{1-p_z}\, \exp{i\xi} \end{pmatrix},
\end{gather*}
generates a unit vector via
\begin{equation}
    \chi^{\dag}(\vec{\sigma}/2)\chi = (\sqrt{1-p_z^2}\,\cos \xi, \sqrt{1-p_z^2}\,\sin \xi, p_z).
\label{SpinorVector}
\end{equation}
The $\sigma$ are the usual Pauli matrices.  Conversely,
the unit vector determines the spinor, up to an
overall arbitrary phase.

    We determine the spinors by demanding that
they reproduce the correct direction for the angular
momentum vector at each vertex.  Normally the source and
target spinors live at different vertices; here they live
at the same vertex because of the $\rS_1$ topology in the
transverse directions.  However,
angular momentum experiences no change when traveling
along the z axis from vertex n-1 to vertex n, since the z holonomy
is a unit matrix.  Further, vertex n will need information about
spinors at vertex n-1, parallel transported to vertex n, in order to
construct covariant differences.  We therefore take the source spinor to
correspond to angular momentum at vertex n-1.
\begin{eqnarray}
    s& = &u(-\beta + \pi/2, \alpha(n-1)/2,+\beta -\pi/2)s(0) := u(n-1)s(0);\nn
    t &=& u(-\beta + \pi/2, \alpha(n)/2,+\beta -\pi/2)s(0) := u(n) s(0).
\label{Peakst}
\end{eqnarray}
Here u is the spin 1/2 representation of the peak holonomy; s
its arguments are the Euler angles.  u is
a rotation through $\alpha/2$ around an axis in the XY
plane making an angle $\beta $ with the X axis.  s(0) corresponds
to the "unrotated" vector, denoted $\hat{p}$ in the text; s corresponds
to the "rotated" vector, denoted by $\hat{P}$.
The latter vector gives the direction of angular momentum.
$\hat{p} = (\cos(\mu + \beta), \sin(\mu + \beta),0)$  happens to be independent of n
(neglecting terms of order $a^2$ in the small amplitude a), therefore s(0) is independent of n.
\begin{gather}
    s(0) =  \begin{pmatrix} 1 \\  \exp{i(\beta + \mu)} \end{pmatrix},
\end{gather}
The spinors of \eq{Peakst} give the correct directions for angular
momenta, for example
\begin{eqnarray}
    t^{\dag}(\sigma_A/2) t &= s(0)^{\dag} u(n)\dag (\sigma_A/2)u(n)s(0)\nn
                    &= s(0)^{\dag}(\sigma_B/2) s(0) D^{(1)}_{BA}\nn
                    &= \hat{p}_B D^{(1)}_{BA} = \hat{P}.
\end{eqnarray}
The last line is the first line of \eq{peakEh}.  For simplicity
I have been computing with unit spinors, but strictly speaking
s and t should be multiplied by $\sqrt[4]{j(j+1)}$ to give the
angular momentum vectors the correct length.

    The spinor t in full detail is
\[
    t  = \begin{bmatrix} \cos(\alpha/4) + i \sin(\alpha/4)\exp(i\mu) \\
               \cos(\alpha/4) \exp[i(\beta + \mu)] + i \sin(\alpha/4)\exp(i\beta)
    \end{bmatrix},
\]
which shows that the key U(N) variables (spinors now, not holonomies)
vary sinusoidally with n.  In the general case the spinors produce
an SU(2) result.   In the planar case, the spinors must generate an
O(3) symmetry; hence the half-of-half-angle cosines and sines in the
j = 1/2 case.

    The U(N) formalism contains a fundamental holonomy operator which is
not simply related to the x and y holonomies used in the present  paper.
That fundamental operator contains a holomorphic part, one which depends
only on unstarred spinors, and therefore has a peak value depending only on
unstarred spinors s and t.
The holonomies used in the present paper must be constructed using both starred and
unstarred spinors.  For example, the eigenspinors and eigenvectors $\exp(\pm i\alpha/4)$
of u can be used to construct u.
\begin{eqnarray}
    u(n) &=& \chi \exp(i \alpha/4) \chi^{\dag} +  C \chi^* \exp(-i \alpha/4)(C \chi^*)^{\dag};\nn
    \chi &=& \chi(0,\beta).
\label{UsualHolonomy}
\end{eqnarray}
C is the usual charge conjugation matrix $-i \sigma_Y$;
$\chi$ and C$ \chi^*$ form a complete set.   In the U(N)
formalism, the matrix \eq{UsualHolonomy} cannot qualify as fundamental, because
both terms on the right contain starred spinors.

    The following is an example of a variable which is holonomic,
therefore plays a fundamental role in the U(N) formalism.
This variable is especially simple to compute, because in the planar
case every transverse x holonomy has the same axis of rotation
(and similarly for the y holonomies).
\begin{eqnarray}
    \rF[t,s] &:=& [C t^*]^{\dag}s \nn
            & =& [C u^*(n) s(0)^*]^{\dag}u(n-1)s(0) \nn
            & =& s(0)^{\mbox{tr}}C^{\dag} u(-\beta + \pi/2, [\alpha(n-1) - \alpha(n)]/2,+\beta -\pi/2)s(0)\nn
            &=& 2 \exp[i(\beta + \mu)][ \sin(\Delta)\sin(\mu)]. \nn
    \Delta &=& [\alpha(n) - \alpha(n-1)]/4.
\end{eqnarray}
The overall phase can be removed by changing the arbitrary
overall phases of the basic spinors.

    Note all boosts have been fixed when SL(2,C) is
reduced to SU(2) in the canonical approach.  Ordinarily one would construct the
intertwiners at each vertex from
F[i,j] variables,
because those are
SL(2,C) invariant as well as SU(2) invariant.  However,
when boosts are fixed, one may use also E[i,j] variables,
which are only SU(2) invariant.
\begin{eqnarray}
    \rE[t,s] &:=& t^{\dag}s\nn
        &=& 2 \cos(\Delta) +2i \sin(\Delta) \cos(\mu).
\end{eqnarray}
In the z direction one must use E, since all z angles are zero
and the corresponding F[t,s] vanishes.

    The formalism developed in this paper uses a matrix
H which lies in the complex extension of SU(2).  The
columns of this matrix also occur in the U(N) formalism.
The spin 1/2 representation of H = $\exp[ \vec{p}\cdot \vec{\rS}]$
is
\[
    H^{(1/2)} = [\exp(p/2)/2] \begin{bmatrix} 1& \exp[-i(\beta + \mu)]\\
                                        \exp[i(\beta + \mu)] &1
                            \end{bmatrix} ,
 \]
in the limit of moderately large p $\sim$ 5.  The two columns of
H are essentially the spinor s(0).

    The U(N) formalism is slightly less intuitive than the
usual formalism, because the spinor is less intuitive than
the associated vector.  However, the usual formalism works
directly with the vector, and is not particularly
intuitive either.  A major motivation for the present paper
was to build some intuition.

    Turning from intuition to computation:
when one considers states of spin higher
than 1/2, the U(N) expressions are easier to manipulate.
One encounters factorials which are actually explicit
expressions for Clebsch-Gordan coefficients.  Given these expressions, usually
it is easy to recouple without consulting a table of 3J symbols.

\end{document}